\documentclass[12pt]{article}
\usepackage[T1]{fontenc}
\usepackage{bm}
\usepackage{cite}
\usepackage{tikz}
\usepackage{mathrsfs}
\usepackage{slashed}
\usepackage{graphicx} 
\usepackage{hyperref}
\usepackage{verbatim}
\usepackage{color}
\usepackage{enumerate}
\usepackage{authblk}
\usepackage[hang,flushmargin]{footmisc}
\usepackage{lipsum}
\usepackage[font=footnotesize]{caption}
\usepackage[utf8]{} 
\usepackage{empheq}
\usepackage{amssymb,amsmath,amsfonts}
\usepackage[makeroom]{cancel}
\usepackage{hyperref}
\usepackage{color}

﻿
﻿
\numberwithin{equation}{section}

\newtheorem{rem}{\textit{Remark}}[section]

\usepackage{physics}

﻿
\definecolor{blue-violet}{rgb}{0.54, 0.17, 0.89}
\definecolor{PineGreen}{cmyk}{0.92, 0, 0.59, 0.25}
\definecolor{OliveGreen}{cmyk}{0.64, 0, 0.95, 0.40}
\definecolor{RawSienna}{cmyk}{0, 0.72, 1, 0.45}
\definecolor{Gray}{cmyk}{0, 0, 0, 0.50}
\definecolor{MidnightBlue}{cmyk}{0.98, 0.13, 0, 0.43}
\definecolor{Orange}{cmyk}{0, 0.61, 0.87, 0}
\definecolor{LimeGreen}{cmyk}{0.50, 0, 1, 0}
\definecolor{Green}{cmyk}{1, 0, 1, 0}

\newcommand{\adstwo}{\mathrm{AdS}_2}
\newcommand{\cL}{\mathcal{L}}
\newcommand{\Lgr}{\mathscr{L}}﻿
\newcommand{\SLtwoR}{\mathrm{SL}(2,\mathbb{R})}
\newcommand{\Uone}{\mathrm{U}(1)}
\newcommand{\sltwoR}{\mathfrak{sl}(2,\mathbb{R})}
\newcommand{\uone}{\mathfrak{u}(1)}

﻿

﻿
﻿
﻿

﻿\newcommand{\Lc}{\mathcal{L}}
\newcommand{\eps}{\epsilon}
﻿
\renewcommand{\tilde}{\widetilde}

\usepackage{mathrsfs}

﻿

﻿
﻿
\begin{document}
﻿
\title{\!\!\LARGE{Asymptotic stability of axisymmetric perturbations of near-horizon extremal Kerr with a Schwarzian-deformed throat}\vspace{5pt}}
﻿
\author[a]{\normalsize{Marcela C\'ardenas}\footnote{\href{mailto:marcela.cardenasl@uss.cl}{marcela.cardenasl@uss.cl}}}
\author[b]{\normalsize{Benjam\'in Cuevas-Tapia}\footnote{\href{mailto:benjamin.cuevas@uc.cl}{benjamin.cuevas@uc.cl}}}
\author[c]{\normalsize{Claudio Mu\~noz}\footnote{\href{mailto:cmunoz@dim.uchile.cl}{cmunoz@dim.uchile.cl}}}
\author[d]{\normalsize{Jessica Trespalacios\footnote{\href{mailto:jessica.trespalacios@udec.cl}{jessica.trespalacios@udec.cl}}}}
﻿\author[c]{\normalsize{Hanne Van Den Bosch}\footnote{\href{mailto:hvdbosch@dim.uchile.cl}{hvdbosch@dim.uchile.cl}}}
\affil[a]{\footnotesize\textit{Universidad San Sebasti\'an, Facultad de Ingenier\'ia,  Bellavista 7, Recoleta, Santiago, Chile.}}
\affil[b]{\footnotesize\textit{Facultad de F\'isica, Pontificia Universidad Cat\'olica de Chile, Vicu\~na Mackenna 4860, Macul, Santiago, Chile.}}
\affil[c]{\footnotesize\textit{CNRS and Departamento de Ingenier\'{\i}a Matem\'atica and Centro de Modelamiento Matem\'atico (CNRS IRL2807), Universidad de Chile, Casilla 170 Correo 3, Santiago, Chile}}
\affil[d]{\footnotesize\textit{Departamento de Ingenier\'ia Matem\'atica, Facultad de Ciencias F\'isicas y Matem\'aticas, Universidad de Concepci\'on, Concepci\'on, Chile.}}
\date{}
﻿
\maketitle
﻿

\begin{abstract}
We study the asymptotic stability of axisymmetric gravitational perturbations
of the near-horizon geometries of (near-)extremal Kerr black holes and consider
the family of backgrounds in which the $\adstwo$ factor of NHEK is deformed by a
Schwarzian mode. Requiring the Killing parameter to be local in the
background makes the Gelfand--Dikii polynomials natural solutions of the
Killing equation, and that locality requirement brings stationary solutions of the Korteweg-de Vries equation. The hierarchy selects a family of backgrounds, which remain $\sltwoR\oplus\uone$ throughout them: the
first level gives NHEK and near-NHEK, the second, solitons that depart from one
of these geometries and return to it. We linearize Einstein's equations around
this family, where the axisymmetric sector reduces to a wave equation on the
nearly-$\adstwo$ base, and analyze this equation by energy and virial methods.
For NHEK we obtain a conserved energy, an integrated local energy estimate and
asymptotic stability in a weighted norm; for a nonconstant Schwarzian the
energy becomes a Lyapunov functional whose monotonicity is dictated by the sign
of its derivative, yielding uniform bounds for
integrable ones, and mild local decay along the KdV soliton. We also find a monotone horizon-localized norm, 
which is not of Aretakis type, showing that stability properties are directly sensitive to the
near-horizon structure encoded in the Schwarzian background. We compute the associated Regge--Teitelboim
charges at a late-time corner of the throat using a radial ADM foliation of the
spacetime.

\end{abstract}

\tableofcontents
﻿

\section{Introduction}\label{sec:intro}

The question of whether the Kerr family of black holes is stable under
gravitational perturbations has organized much of mathematical relativity
since Teukolsky's decoupling of the linearized Einstein equations into a
single master equation for the extreme-helicity Weyl scalars
\cite{Teukolsky:1972my,Teukolsky:1973ha}. For subextremal Kerr the picture is
by now well developed: mode stability holds, and quantitative decay estimates
for solutions of the wave and Teukolsky equations have been established by a
combination of vector field, energy and integrated local energy decay methods.
The extremal case $a=M$ is markedly different. Aretakis showed
\cite{Aretakis} that transverse derivatives of a scalar field do not decay
along the event horizon of an extremal black hole and that, at sufficiently
high order, they grow without bound; the same instability was subsequently
shown to afflict linearized gravitational and electromagnetic perturbations of
extreme Kerr \cite{Lucietti:2012sf}, and to develop even from data supported
arbitrarily far from the horizon \cite{Aretakis:2012bm}. In the axisymmetric
sector, which is the one we shall work in, the wave equation on extreme Kerr
admits energy estimates that give boundedness and power-law decay up to and
including the horizon, while leaving the transversal derivatives
uncontrolled \cite{Aretakis:2011hc}. This horizon instability is not detected
by the mode analysis, coexists with decay in the exterior, and is tied to the
degeneracy of the horizon rather than to any particular matter content.

Two developments make it natural to attribute the effect to the near-horizon
region rather than to the extremal black hole as a whole. At near-extremality,
Gralla, Zimmerman and Zimmerman \cite{Gralla:2016sxp} found that scalar,
electromagnetic and gravitational perturbations of a rapidly rotating Kerr
black hole produce a transient growth of the energy density and of the tidal
forces measured by infalling observers, lasting arbitrarily long as
extremality is approached, and attributed it to the near-horizon geometry.
More directly, Gralla and Zimmerman \cite{Gralla:2018xzo} showed that the
decay rates and the growth of high-order transverse derivatives follow from
the emergent near-horizon conformal symmetry alone, through the
$\adstwo$ structure of the throat.

The near-horizon region of extremal Kerr is captured by the NHEK geometry of
Bardeen and Horowitz \cite{Bardeen:1999px}, obtained from Kerr by a
corotating scaling limit. NHEK is not asymptotically flat: it is a warped
fibration of a $\Uone$ over an $\adstwo$ base, and the two Killing
vectors of Kerr are enhanced to an $\SLtwoR\times\Uone$ isometry
algebra. This enhancement underlies the Kerr/CFT correspondence
\cite{Guica:2008mu} and ties the throat to the holography of two-dimensional
dilaton gravity, whose study goes back to the $\adstwo$ fragmentation of
extremal throats \cite{Maldacena:1998uz}. In a nearly-$\adstwo$
spacetime the reparametrization symmetry of the boundary time is broken both
spontaneously and explicitly, and the resulting soft sector is governed by a
Schwarzian action
\cite{Almheiri:2014cka,Jensen:2016pah,Maldacena:2016upp,Engelsoy:2016xyb}.
That this sector controls the low-temperature dynamics of near-extremal black
holes has been established for charged solutions
\cite{Nayak:2018qej,Iliesiu:2020qvm} and for rotating ones
\cite{Moitra:2019bub}, while the Schwarzian mode of the extremal Kerr throat
in particular is the object studied in
\cite{Kapec:2019hro,Castro:2019crn}. Linearized gravitational
perturbations of NHEK have been analyzed in
\cite{Amsel:2009ev,Dias:2009ex}, and it was shown in
\cite{Dias:2009ex} that NHEK is stable at the level of modes. Set against the
Aretakis result, this raises the question posed sharply by Durkee and Reall
\cite{Durkee:2010ea}: to what extent does the stability, or instability, of a
near-horizon geometry control that of the parent black hole? Because the
near-horizon geometry does not share the asymptotic regions of the full
solution, the answer cannot be read off from decay at null infinity, and the
statement one proves depends delicately on the norm, the foliation and the
boundary conditions employed.

\medskip 

The present work addresses a version of this question in which the background
itself is allowed to vary. Rather than treating NHEK and near-NHEK as two
isolated geometries, we work with the family of solutions
\cite{Kapec:2019hro,Castro:2019crn} obtained by acting on NHEK with the
improper coordinate transformation generated by a reparametrization $f(t)$,
whose only imprint on the metric is through the Schwarzian derivative. It makes manifest
the conformal symmetry of the throat and the different scaling limits that
resolve it at different depths, and it converts the comparison between
backgrounds into the study of a single function,the Schwarzian, entering in the
perturbation equation as a coefficient. That such a family exists at all is
worth a remark, since it was argued in \cite{Amsel:2009ev} that every solution
asymptotic to NHEK is diffeomorphic to NHEK itself, so that the throat carries
no nontrivial dynamics.

 Revisiting the symmetries of the
family, we realize that the Schwarzian $\mathcal{L}=\{f(t),t\}$ is not free. We allow the gauge parameter to be field dependent,
$\epsilon=\epsilon[\mathcal{L}]$, while requiring it to remain local in the
background, a differential polynomial in $\mathcal{L}$ and its derivatives.
This is natural once one notices that the transformation law of $\mathcal{L}$
under a reparametrization is generated by the symplectic operator of the
Korteweg--de Vries hierarchy, so that the Gelfand--Dikii polynomials organize the Killing equation level by level.
The first level returns NHEK and near-NHEK; the
second is the stationary KdV equation, whose solitons are pulses that depart
from one of these geometries and return to it. The criterion is purely
kinematic, locality of the Killing parameter, and it constrains the
Schwarzian mode rather than generating a flow for it, singling out a
distinguished set of backgrounds on which to test stability. The role of KdV in
deforming the geometry has been noted before in three dimensions
\cite{Perez:2016vqo,Dymarsky:2020tjh, Cardenas:2021vwo,Cardenas:2025qqi} and in nearly-$\adstwo$, where the throat is deformed by
the KdV charges \cite{Cardenas:2024hah}. Most recently, this has been extended to higher dimensions, where genuinely
four-dimensional perturbations were constructed, carrying the KdV dynamics
and modifying the thermodynamics \cite{Cardenas:2026}.
\medskip

We then study linearized gravitational perturbations of this family of metrics. We adopt the axisymmetric ansatz of
\cite{Castro:2021csm}, which for a constant Schwarzian mode is related by a
near-horizon limit to the Teukolsky master equation, but we retain
the Schwarzian as a background field.

The analysis of this equation is carried out by energy methods rather than by
mode expansion, in the spirit of the vector field method introduced by
Klainerman and of the virial estimates that describe decay in settings where
scattering theory is unavailable \cite{Virial1,Virial2}. These tools have been used to control the long-time
dynamics of solitons of the Einstein equations on Belinski--Zakharov
backgrounds \cite{MT} . 
The advantage here is that the conclusions can be tracked
directly as functions of the background profile $\mathcal{L}$. Taken together, the results give a concrete answer to the question raised
above, at least within the axisymmetric sector: the stability statement one
obtains is not a property of the near-horizon region as such, but of the
background profile of the Schwarzian mode and of the norm in which decay is
measured. The same geometry that is asymptotically stable in the weighted energy norm carries conserved and growing transverse derivatives on its degenerate horizon,
while the thermal deformation damps them; and moving along the KdV family from NHEK to near-NHEK changes a conservation law into a Lyapunov
inequality.

\medskip

 Finally, we compute the conserved charges associated with the
symmetries.  The charges of near-horizon extremal geometries have been studied extensively,
both as the asymptotic Virasoro charges of the Kerr/CFT construction
\cite{Guica:2008mu,Compere:2012jk} and as symplectic charges on the phase
space generated by finite deformations of the throat
\cite{Hajian:2013lna,Compere:2015mza,Compere:2015bea}. Our setting differs
from the latter in that the deformation acts along the time direction rather
than the angular one, so that the charges depend on the surface at which they
are evaluated.
 Since the backgrounds are non-stationary, we use a radial
Regge--Teitelboim foliation \cite{Regge:1974zd,Banados:2004zt} with leaves of
Lorentzian signature and evaluate the charge variation at a late-time corner
of the throat. The soliton backgrounds asymptoting to NHEK and to near-NHEK
carry different charges, the difference being controlled by the amplitude
$\omega$ of the pulse and by the residual temperature $\kappa$.

The paper is organized as follows. Section~\ref{Limits} reviews the Kerr
solution and the scaling limits to NHEK and near-NHEK, and
Section~\ref{Extensions} develops the field-dependent extension of the
symmetry algebra and its relation to the KdV hierarchy.
Section~\ref{sec:perturbations} sets up the axisymmetric gravitational
perturbations, establishes the Petrov type and the gauge invariance of
$\chi$, and derives the wave equation. It contains the energy and virial
estimates for NHEK, for general $\mathcal{L}$ and for the KdV soliton, and
Section~\ref{sec:aretakis} the horizon instability.
Section~\ref{sec:charges} constructs the conserved charges, and
Section~\ref{sec:conclusions} collects our conclusions and open questions.
Three appendices contain, respectively, the reduction to Teukolsky's master
equation, the proofs of the energy estimates, and the conformal diagrams of
the twisted fibration.

﻿

﻿
\section{Kerr black hole and the limits to extremality} \label{Limits}
 We consider the Kerr family of stationary black holes in 4 dimensions, that in natural units $G=c=1$  and using the Boyer–Lindquist coordinates is given by
 \begin{equation}\label{Kerr-metric}
 \begin{aligned}
     ds^{2}=&-\frac{\Sigma \Delta}{(\tilde{r}^{2}+a^2)^2-\Delta a^2 \sin^{2}(\theta)} d\tilde{t}^2+\Sigma \left(\frac{d\tilde{r}^{2}}{\Delta}+d\theta^{2}\right)\\
     &+\frac{\sin(\theta)^2}{\Sigma}\left((\tilde{r}^2+a^2)^{2}-\Delta a^{2}\sin^{2}(\theta)\right)\left(d\tilde{\phi}-\frac{2aM\tilde{r}}{(\tilde{r}^{2}+a^2)^2-\Delta a^2 \sin^{2}(\theta)} d\tilde{t}\right)^2,
 \end{aligned}
 \end{equation}
 where
 \begin{equation*}
\Sigma={\tilde{r}^2}+a^2\cos^{2} (\theta),\qquad\Delta({\tilde r})=(\tilde{r}-r_{-})(\tilde{r}-r_{+}),\qquad r_{\pm}=M\pm\sqrt{M^2-a^2}.
\end{equation*}
The solution is parametrized by its mass $M$ and its angular momentum $J$  by means of the rotation parameter $a=J/M$,  generated by the Killing symmetries $\partial_{t}$ and $\partial_{\phi}$ respectively.  The inner and outer horizons are located at $r_{-}$ and $r_{+}$,  the angular velocity of the horizon $\Omega=a/2Mr_{+}$.  Its temperature is given by
\begin{equation}\label{Hawking-temperature}
 T=(r_+-r_{-})/8\pi M r_{+}.
\end{equation}
  If $a>M$,  there is no event horizon as the Kerr metric describes a naked singularity and if $a=M$, the black hole is said to be extremal as it is rotating at its maximum angular momentum and its temperature vanishes.  We will place ourselves in this particular case and take the co-rotating change of coordinates,
\begin{equation}\label{coordinate-change1}
    \tilde{r}=M+\lambda R, \qquad
    \tilde{t}=\frac{2M^{2}}{\lambda}\,T, \qquad
    \tilde{\phi}=\Phi+\frac{M}{\lambda}\,T .
\end{equation}
In the limit $\lambda\rightarrow 0$,  holding $(T,R,\theta,\Phi)$ fixed,  one obtains the so-called Near-Horizon Extreme Kerr (NHEK) metric \cite{Bardeen:1999px},
\begin{equation}\label{NHEK}
    ds^2= \Lambda(\theta)\left[-R^2 dT^2+\frac{dR^2}{R^2}+d\theta^2\right]+\Gamma(\theta)\left[d\Phi+RdT\right]^2 ,
\end{equation}
for 
\begin{equation}\label{angular-functions}
\Lambda(\theta)=M^{2}(1+\cos^{2}(\theta))\qquad\text{and}\qquad \Gamma(\theta)=\frac{4M^{2}\sin^{2}(\theta)}{(1+\cos^{2}(\theta))}.
\end{equation}
The geometry \eqref{NHEK} exhibits an $\adstwo$ factor associated to the $T-R$ plane and axial symmetry,  related with the metric coefficients being independent on $\Phi$. Surfaces of constant $\theta$ are warped AdS$_{3}$ geometries \cite{Anninos:2008fx}. This metric has an enhanced $\SLtwoR$ symmetry with Killing vectors
\begin{equation}\label{Killin-Sl2R}
\zeta =  \zeta(\epsilon) = \qty(\epsilon(T) + \frac{\epsilon''(T)}{2R^2})\partial_T - R\epsilon'(T)\partial_R
- \frac{\epsilon''(T)}{ R }\partial_\Phi + c_\Phi\partial_\Phi,
\end{equation}

Here $c_\Phi$ is an arbitrary constant associated to the $\Uone$ symmetry and the gauge parameter obeys the condition $\epsilon(T)'''=0\rightarrow \epsilon(T)=c_{-1}T^{2}+c_{0}T+c_{+1} $ where $c_{n} $ are integration constants labeling the generators $L_{n} $ of the $\sltwoR$ algebra,
\begin{equation}
\left[L_{n} ,L_{m}\right] = (n-m)L_{n+m}\qquad\text{for}\qquad{n,m=-1,0,1}.
\end{equation}
In particular,  $c_{+1}$ is connected to the time translation symmetry, and
\begin{equation}\label{SL(2,R)-generators}
L_{-1}=\left(\frac{1}{R ^2}+T^2 \right)\partial_{T}-2  TR\partial_{R} -\frac{2}{R }\partial_{\Phi} \quad, \quad  L_{0}=T\partial_{T}-R\partial_{R} \quad, \quad L_{1}=\partial_{T}\,.
\end{equation}
They close the Lie algebra, 
\begin{equation}\label{Lie-bracket}
\{\zeta[\epsilon_1],\zeta[\epsilon_2]\}^{\mu} = \zeta^{\mu}[\epsilon_3]
\qquad \text{for}\qquad  \epsilon_{3}=\epsilon_1\epsilon_2'-\epsilon_2\epsilon_1'.
\end{equation}
\subsection{Near-NHEK, extensions of the symmetry group and KdV}\label{Extensions}
To explore enhancements of the global conformal isometry group $\SLtwoR$,  we review the study of the symmetries of nearly-$\adstwo$ geometries, as they are responsible for driving black holes away from the extremal zone.  These metrics have been studied extensively, as they represent a closer zoom into the horizon, and are recovered by taking small deviations from extremality \cite{Bredberg:2009pv}, which captures several possible coordinate re-scalings, labelling different deviations that characterize time-like orbits. Specifically, the near-NHEK is obtained by applying the coordinate change
\begin{equation}\label{near_NHEK_coord}
    \tilde{r}=\frac{r_++r_{-}}{2}+\lambda \left( r+\frac{\kappa^2}{4 r} \right), \quad \tilde{t}= 2M^{2}\frac{t}{\lambda}, \quad \phi=\Phi+\frac{M\,t}{\lambda}.
\end{equation}
on the  Kerr metric \eqref{Kerr-metric}. With the additional condition that $r_{\pm}=M\pm \lambda \kappa$, so that $a^{2}=r_{+}r_{-}=M^{2}-\lambda^{2}\kappa^{2}$, the decoupling limit $\lambda\rightarrow0$ gives 
\begin{equation}\label{4near-NHEK}
ds^2 = \Lambda(\theta)\left[-\qty(r -\frac{\kappa ^2}{4r})^2\dd t^2 + \frac{\dd r^2}{r^2}+ \dd\theta^2\right] + \Gamma(\theta)\left[\dd\phi + \qty(r+\frac{\kappa ^2}{4r})\dd t\right]^2.
\end{equation}
Similarly to the previous limit, it yields a metric that does not depend on $\lambda$, but does depend on $\kappa$ which captures a power law deviation from extremality of the Hawking temperature \eqref{Hawking-temperature} as,
\begin{equation}
 T=\frac{\lambda  \kappa }{4\pi M^2}-\frac{\lambda ^2\kappa ^2}{4\pi M^3}+\mathcal{O}(\lambda ^3\kappa ^3).
\end{equation}

To study the previous metrics in a more general setup, we consider a set of geometries proposed in \cite{Kapec:2019hro} and revisited in \cite{Castro:2019crn}, that acknowledges the existence of several possible scalings while approaching the near horizon region of an extremal black hole. We take the improper coordinate transformation \footnote{This coordinate change is actually inspired from three dimensions \cite{Roberts:2012aq}.} ,   
\begin{equation}\label{nonlinear coord-change}
    T=f(t)+\frac{2\left( f'\right)^2\, f''}{4 r^{2}\,f^2-f''},\qquad
    R=\frac{r}{f'}-\frac{f''}{4 r\,\left( f' \right)^3},\qquad
    \Phi=\phi+\log\left[\frac{2rf'-f''}{2 rf'+f''}\right],
\end{equation}
where $f(t)$ is the mode function associated to the time reparametrization symmetry.  Here and throughout the paper we use the convention $\partial_{t}=(\,)'$ for functions depending on the $t$-coordinate only. The change of coordinates \eqref{nonlinear coord-change} is applied in \cite{Castro:2019crn} and one obtains
\begin{equation}\label{4dmetric}
ds^2 = \Lambda(\theta)\qty(-\qty(r + \frac{\mathcal{L}(t)}{2r})^2\dd t^2 + \frac{\dd r^2}{r^2}+ \dd\theta^2) + \Gamma(\theta)\qty(\dd\phi + \qty(r - \frac{\mathcal{L}(t)}{2r})\dd t)^2,
\end{equation}
with
\begin{equation}\label{Schwarzian}
 \mathcal{L}(t)= \left\{f,t\right\}=\frac{f'''}{f'}- \frac{3}{2}\left(\frac{f''}{f'}\right)^{2}.
\end{equation}
This is the Schwarzian derivative $\left\{f,t\right\}$, which is invariant function under the action of the $\SLtwoR$ transformation
 \[
f(t)\to \tilde{f}(t)=\frac{af(t)+b}{cf(t)+d}\quad \mbox{with} \quad ad-bc=1.\]

\medskip

This metric contains the NHEK geometry by the choice $f(t)=t$, giving $\cL = 0$, and the near-NHEK metric by the choice $f(t)=\exp (\kappa t)$, and crucially,  makes manifest the conformal symmetry. Indeed, computing the group symmetries generated by the vector $\zeta^{\mu}=\{\zeta^{t},\zeta^{r},\zeta^{\theta},\zeta^{\phi }\}$,  that preserves the form of the metric \eqref{4dmetric}, 
\begin{equation}\label{gauge-transformation}
\delta g_{\mu\nu} = \mathcal{L}_\zeta g_{\mu\nu},
\end{equation}
one finds
\begin{equation}\label{general-Killin}
\zeta = \qty(\epsilon(t) + \frac{\epsilon''(t)}{\qty(2r^2 + \mathcal{L}(t))})\partial_t - r\epsilon'(t)\partial_r 
- \frac{2r\epsilon''(t)}{2 r^2 + \mathcal{L}(t)}\partial_\phi + c_\phi\partial_\phi,
\end{equation}
where $\epsilon$ is an arbitrary function of time related with the time reparametrization symmetry and $c_\phi$ is an arbitrary constant associated to rotational invariance. From \eqref{gauge-transformation}, we also obtain the additional symmetry transformation \footnote{ Equation (2.18) is the transformation law of a CFT$_2$ stress tensor of weight two, the term $\epsilon'''$ playing the role of the central extension. Writing $T=\alpha\mathcal{L}$ brings it to the standard form with $c=12\alpha$.}, 
\begin{equation} \label{transformation}
\delta \mathcal{L}= \epsilon(t) \mathcal{L}'(t) + 2\epsilon'(t) \mathcal{L}(t) + \epsilon'''(t).
\end{equation}

\subsubsection{Solving the Killing equation}

An isometry of \eqref{4dmetric} is a vector \eqref{general-Killin} whose
action leaves the Schwarzian mode inert, $\delta_\epsilon\mathcal{L}=0$. By
\eqref{transformation} this is the linear third-order equation
\begin{equation}\label{Killing-condition}
\mathcal{D}_1\epsilon \;\equiv\;
\epsilon''' + 2\mathcal{L}\epsilon' + \mathcal{L}'\epsilon \;=\; 0 ,
\end{equation}
is the classical third-order equation
associated with the Hill operator.  Indeed,  if $\psi_1,\psi_2$ are any two solutions of
\begin{equation}\label{Hill}
\psi'' + \tfrac12\,\mathcal{L}\,\psi = 0 ,
\end{equation}
then each of the three products $\psi_1^2$, $\psi_1\psi_2$, $\psi_2^2$ solves
\eqref{Killing-condition}, as follows by direct substitution, and these span
its solution space. Since \eqref{Hill} is of second order, the space of
isometry parameters is $\mathrm{Sym}^2$ of a two-dimensional space, hence
three-dimensional for \emph{every} profile $\mathcal{L}$. Together with
$c_\phi\partial_\phi$ this is the algebra
\begin{equation}\label{isometry-algebra}
\sltwoR\oplus\uone,
\end{equation}
the composite parameter of \eqref{Lie-bracket} being the symmetric-square
representation.  Then, every member of the family
\eqref{4dmetric} is obtained from NHEK by the coordinate change
\eqref{nonlinear coord-change},  being all locally isometric where no
choice of $f$ can enlarge or reduce the algebra.

What does change with the background is the realization. For $\mathcal{L}=0$
one has $\psi\in\{1,t\}$ and the familiar parameters $\epsilon\in\{1,t,t^2\}$
of \eqref{Killin-Sl2R}. For near-NHEK, where $f=e^{\kappa t}$ gives
$\mathcal{L}=\{e^{\kappa t},t\}=-\kappa^2/2$, the Hill equation has
$\psi\in\{e^{\pm\kappa t/2}\}$ and the parameters are
$\epsilon\in\{1,\,e^{\kappa t},\,e^{-\kappa t}\}$, where the real exponentials appear because $\mathcal{L}<0$.

Indeed,  the sign of $\mathcal{L}$ is geometrically meaningful. The metric factor $r+\mathcal{L}/2r$ of
\eqref{4dmetric} vanishes at $2r^2=-\mathcal{L}$, so for $\mathcal{L}<0$ a
Killing horizon sits at finite radius, at $r=\kappa/2$ in the near-NHEK case,
whereas for $\mathcal{L}\ge 0$ the coordinate $r$ ranges over all of
$(0,\infty)$ without encountering one. This distinction reappears in
Section~\ref{sec:perturbations}, where the weight
$(\mathcal{L}+2r^2)/r^2$ controlling the energy degenerates precisely at that
radius.  In this case,  $\mathcal{L}$ acts as deformation of the background, where this assumption will act favourably in the study of the asymptotic stability of axisymmetric perturbations.

\subsubsection{Locality as a selection principle and Korteweg--de Vries}
The question we answer now is 
for which profiles of $\mathcal{L}$,  the isometry parameter $\epsilon$ is expressible
locally in terms of the background, that is, as a differential
polynomial $\epsilon=\epsilon[\mathcal{L},\mathcal{L}',\dots]$. Imposing locality
converts \eqref{Killing-condition} from an ordinary differential equation for
$\epsilon$ on a fixed background into a condition on $\mathcal{L}$ itself.  To start, we notice that the Korteweg--de Vries hierarchy is the structure that organizes the local
solutions. The transformation law \eqref{transformation} is generated by the
first Hamiltonian operator of KdV,
$\mathcal{D}_1=2\mathcal{L}\partial_t+\mathcal{L}'+\partial_t^3$ \cite{Gervais:1985fd}, and
together with the second structure $\mathcal{D}_2=\partial_t$ it defines the
Lenard recursion
\begin{equation}\label{recurrence}
\mathcal{D}_2\,\epsilon_{l+1} = \mathcal{D}_1\,\epsilon_{l},
\qquad \epsilon_0 = 1 ,
\end{equation}
whose solvability within differential polynomials is the statement that
$\mathcal{D}_1\epsilon_l$ is a total time derivative. The resulting
Gelfand--Dikii polynomials\footnote{That these exhaust the local solutions of
\eqref{Killing-condition} is the standard Gelfand--Dikii result; we use only
that they provide an infinite family of them.} are
\begin{equation}\label{GD-polynomials}
\epsilon_0 = 1 ,\qquad
\epsilon_1 = \mathcal{L} + \bar\epsilon_1 ,\qquad
\epsilon_2 = \tfrac32\mathcal{L}^2 + \mathcal{L}''
+ \bar\epsilon_1\mathcal{L} + \bar\epsilon_2 ,\qquad \dots
\end{equation}
Demanding that $\epsilon_l$ be an isometry parameter,
$\mathcal{D}_1\epsilon_l=0$, then selects a class of backgrounds at each
level. The first level gives the condition
\begin{equation}\label{chiral}
\mathcal{L}' = 0 ,
\end{equation}
whose solutions are exactly NHEK and near-NHEK. The second gives
\begin{equation}\label{KdV}
\mathcal{L}''' + 3\mathcal{L}\mathcal{L}' + \bar\epsilon_1\mathcal{L}' = 0 ,
\end{equation}
the stationary KdV equation, whose solitonic solutions describe a transient
excursion away from extremality,  as it will be discussed in detail along Section~\ref{sec:perturbations} and Section \ref{Corner-charges}.  Each further level yields a higher stationary
equation and hence a further class of admissible profiles.

By the recursion \eqref{recurrence}, the stationary condition\footnote{Here \emph{stationary} is used in the sense of the KdV hierarchy: the
polynomials $\epsilon_l$ generate flows
$\partial_{s}\mathcal{L}=\mathcal{D}_1\epsilon_l$ in an auxiliary parameter
$s$, and \eqref{KdV} is the condition that $\mathcal{L}$ be invariant under
that flow. } at level $l$ can
be stated equivalently as
\begin{equation}\label{G-D-Conservation}
\mathcal{D}_1\epsilon_l = 0
\qquad\Longleftrightarrow\qquad
\epsilon_{l+1}' = 0 ,
\end{equation}
so that the polynomial following the one that generates the isometry is
constant on the corresponding backgrounds. At the first level the statement is
stronger: since $\mathcal{L}'=0$,  every Gelfand--Dikii polynomial
reduces to a constant, and the whole tower is conserved. This is special to
the constant $\mathcal{L}$ case; already at the second level one has
$\epsilon_1'=\mathcal{L}'\neq0$, while $\epsilon_2$ collapses to the constant
$\bar\epsilon_2$.

We emphasize that this is a statement about backgrounds, not about evolution.
In the $1+1$ dimensional hierarchy the polynomials generate flows
$\partial_{s}\mathcal{L}=\mathcal{D}_1\epsilon_l$ in an auxiliary time $s$;
here $\mathcal{L}$ depends on $t$ alone and the geometry is radially
independent, so the flow parameter is absent and only the stationary sector
survives. This is also why the KdV potential is a function of time rather than
of a spatial coordinate. In the Euclidean section, $t=i\tau$ with
$\tau\sim\tau+\beta$, that coordinate becomes a genuine circle and the
hierarchy acquires its usual periodic setting, with $\beta$ the inverse
temperature.

\subsection{Example: a soliton background geometry }

We can now return to \eqref{Killing-condition} on one of the profiles singled
out above and obtain the local generators for this case. Take the soliton of
\eqref{KdV}, $\mathcal{L}=4b^2\,\mathrm{sech}^2(bt)$ with
$\bar\epsilon_1=-4b^2$. The Hill equation \eqref{Hill} reads
$\psi''+2b^2\mathrm{sech}^2(bt)\,\psi=0$ and is solved by
\begin{equation}\label{Hill-soliton}
\psi_1 = \tanh(bt), \qquad
\psi_2 = t\tanh(bt) - \frac1b ,
\end{equation}
the second obtained from the first by reduction of order, with unit Wronskian.
The three isometry parameters are the corresponding products,
\begin{equation}\label{soliton-killing}
\epsilon = c_1\,\tanh^2(bt)
+ c_2\left(t\tanh^2(bt)-\frac{\tanh(bt)}{b}\right)
+ c_3\left(t\tanh(bt)-\frac1b\right)^2 ,
\end{equation}
and $c_1,c_2,c_3$ are the three constants of \eqref{isometry-algebra}. As
$t\to\pm\infty$ the soliton decays and \eqref{soliton-killing} approaches
$c_1+c_2t+c_3t^2$, recovering the NHEK parameters $\{1,t,t^2\}$, as it must
since $\mathcal{L}\to0$ there.
 
The algebra these generate is again $\sltwoR$, and the
products of Hill solutions provide a canonical basis for it. Normalizing the
Wronskian of \eqref{Hill} to unity, $\psi_1\psi_2'-\psi_2\psi_1'=1$, the
composite parameter of \eqref{Lie-bracket} gives
\begin{equation}\label{sl2-Hill-basis}
[\psi_1^2,\,\psi_1\psi_2] = \psi_1^2 , \qquad
[\psi_1^2,\,\psi_2^2] = 2\,\psi_1\psi_2 , \qquad
[\psi_1\psi_2,\,\psi_2^2] = \psi_2^2 ,
\end{equation}
which are the relations $[L_n,L_m]=(n-m)L_{n+m}$ of
\eqref{SL(2,R)-generators} under the identification
$\psi_1^2\leftrightarrow L_1$, $\psi_1\psi_2\leftrightarrow L_0$,
$\psi_2^2\leftrightarrow L_{-1}$. The structure constants in
\eqref{sl2-Hill-basis} involve only the Wronskian, so they are the same for
every profile $\mathcal{L}$: on NHEK the basis is $\{1,t,t^2\}$ and on the
soliton it is the one appearing in \eqref{soliton-killing}.
 
Only the first of the three is local in the background:
$\psi_1^2=1-\mathcal{L}/4b^2$, so that $c_1$ reproduces the Gelfand--Dikii
parameter $\epsilon_1=\mathcal{L}+\bar\epsilon_1$ of \eqref{GD-polynomials} up
to normalization. The other two cannot be written as differential polynomials
in $\mathcal{L}$: any such polynomial is a bounded function of $t$ once
$\mathcal{L}$ and all its derivatives decay, whereas $\psi_1\psi_2$ and
$\psi_2^2$ grow linearly and quadratically.

 The integrability of the Killing
equation under the requirement of locality,  provides a purely kinematic criterion so that it endows the Schwarzian mode
with dynamics and singles out a distinguished set of backgrounds.  We analyze in the remainder of the paper the stability of
the geometries,  with
$\mathcal{L}(t)$ entering the perturbation equation as an explicit
time-dependent coefficient.

\section{Gravitational perturbations}\label{sec:perturbations}

In this section we study axisymmetric gravitational perturbations of the
family \eqref{4dmetric}. We adopt the ansatz of \cite{Castro:2021csm}, which
for constant $\mathcal{L}$, and in particular for $\mathcal{L}=0$,  is
related by a near-horizon limit to the axisymmetric Teukolsky master equation
of the Kerr black hole.  A similar ansatz was proposed in \cite{Mariani:2025hee}. Here, we retain the Schwarzian as a
background field, so that the profiles selected in Section~\ref{Extensions}
enter the perturbation equation as an explicit time-dependent coefficient,
and the stability analysis can be carried out as a function of the
background.

Our main tool is the implementation of vector field methods adapted to the background. This a robust geometric approach (see \cite{Alinhac, Aretakis2012, Dafermos2024} for a complete introduction of the methodology) that allows to say that, for functions that are sufficiently \emph{well-behaved}, an $L^2$-norm with weights decays. In its classical form \cite{Klainerman1985},  vector fields are used as multipliers, contracted with the energy-momentum tensor, and produce energy currents whose divergence is governed by the deformation tensor of the field. When the field is a Killing field, the current is conserved and coincides with the Noether current of the corresponding symmetry; when it is not, the failure of the symmetry appears as an explicit bulk term, which may either spoil conservation or, if it has a definite sign, yield decay.

\medskip 
We first derive the wave-type equation from the linearized Einstein equations for the scalar that quantifies the size of the perturbation and relate it to the Weyl scalars in Section~\ref{sec:axisym}. Then, in Section~\ref{sec:killing-fields} we derive the currents and charges for a general vector field and its relation with the background field $\cL$. We apply these general results to different backgrounds in Section~\ref{sec:vector_fields_aplications}. Section~\ref{sec:virial} specifies to the pure NHEK solution and obtains decay of solutions from a virial argument. Finally, Section~\ref{sec:aretakis} returns to the Near-NHEK family and obtains the divergence of suitable norms near the horizon.

\subsection{Linearized Einstein's equations for axisymmetric perturbations}\label{sec:axisym}

We start by deriving the equations for axisymmetric perturbations of a general near-NHEK solution from the spacetime
\begin{equation} \label{eq:define_pert_metric}
\begin{aligned}
ds^2 = &~{} M^2(1 + \cos^2\theta + \delta\chi(x^{a},\theta))[g_{ab}\dd x^{a}\dd x^{b} + \dd\theta^2] \\
&~{}+ 4M^2\frac{\sin^2\theta}{1 + \cos^2\theta + \delta\chi(x^{a},\theta)}[\dd\phi + A_{a}\dd x^{a}+ \delta\mathcal{T}]^2.
\end{aligned}
\end{equation}
Recall that the near-NHEK metric depends on the background $\mathcal{L}$ through
\begin{equation}\label{2dmetric}
g_{ab}\dd x^a \dd x^b = -\qty(r + \frac{\mathcal{L}(t)}{2r})^2\dd t^2 + \frac{\dd r^2}{r^2}\qquad\text{and}\qquad A_a\dd x^a = \qty(r - \frac{\mathcal{L}(t)}{2r})\dd t.
\end{equation}
The deviation from the background solution is driven by the scalar $\chi$ and the one-form $\mathcal{T}=\mathcal{T}_{a}(x^{a},\theta)dx^{a}$, multiplied by the perturbative constant $\delta$.

Throughout the paper we use the shorthand
\begin{equation}\label{B-def}
B(t,r):=2r^2+\mathcal{L}(t),
\end{equation}
so that the lapse of the $\adstwo$ base is $r+\mathcal{L}/2r=B/2r$ and $g_{tt}=-B^2/4r^2$. We also recall that the only nonvanishing component of the connection in \eqref{2dmetric} is
\begin{equation}\label{At-def}
A_t(t,r)=r-\frac{\mathcal{L}(t)}{2r}.
\end{equation}

\subsubsection{Weyl scalars}
To analyze the nature of these perturbations we use the Newman--Penrose
formalism \cite{Newman:1961qr,chandrasekhar1998mathematical}, in which the
metric is written as
$g_{\mu\nu}=-l_\mu n_\nu-l_\nu n_\mu+m_\mu\bar{m}_\nu+\bar{m}_\mu m_\nu$ in
terms of a complex null tetrad normalized as $-\,l\cdot n=m\cdot\bar{m}=1$ for
signature $(-,+,+,+)$, with all remaining inner products vanishing, and with
$l$ and $n$ tangent to affinely parametrized null geodesics. A tetrad adapted
to \eqref{4dmetric} is, in terms of \eqref{angular-functions}, \eqref{B-def} and \eqref{At-def},
\begin{align}
l&= \frac{2r^2A_t}{\Lambda(\theta)\,B^2}\left(-\partial_t+\frac{B}{2}\,\partial_r+A_t\,\partial_\phi\right), \label{Newman-Penrose-basis} \\
n &= \frac{1}{A_t}\left(-\partial_t-\frac{B}{2}\,\partial_r+A_t\,\partial_\phi\right), \notag\\
m &= \frac{1}{\sqrt{2}}\left(\frac{1}{\sqrt{\Lambda(\theta)}}\,\partial_\theta+\frac{i}{\sqrt{\Gamma(\theta)}}\,\partial_{\phi}\right).\notag
\end{align}
The vectors $l$ and $n$ differ only in the sign of their radial component: both are built on the horizontal combination $-\partial_t+A_t\partial_\phi$, orthogonal to the $\Uone$ fiber, while $m$ is the orthonormal frame of the fiber block $\Lambda\,d\theta^2+\Gamma(d\phi+A_t dt)^2$.
Of the five Weyl scalars built by contracting the Weyl curvature $C_{\mu\nu\alpha\beta}$ with
this basis we shall need the two of helicity $\pm2$,
\begin{equation}\label{eq:weyl-scalars}
\Psi_0 = C_{\mu\nu\alpha\beta}\,l^\mu m^\nu l^\alpha m^\beta,
\qquad
\Psi_4 = C_{\mu\nu\alpha\beta}\,n^\mu \bar{m}^\nu n^\alpha \bar{m}^\beta,
\end{equation}
together with $\Psi_2$. On the unperturbed geometry \eqref{4dmetric} only the
latter survives,
\begin{equation}
\Psi_0=\Psi_1=\Psi_3=\Psi_4=0,
\qquad
\Psi_2=-\frac{i}{M^{2}(\cos\theta-i)^{3}},
\end{equation}
so that \eqref{4dmetric} is of Petrov type D, as is the Kerr geometry from
which it descends. Two standard consequences will be used. By the
Goldberg--Sachs theorem the principal null directions of a vacuum type D
spacetime are shear-free geodesic congruences; and $\Psi_0$ and $\Psi_4$ are
the only irreducible components of the Weyl tensor carrying two transverse
directions and helicity $\pm2$, hence the only ones that propagate. A
perturbation therefore carries gravitational degrees of freedom precisely when
it renders them non-vanishing.

\medskip

At linear order in the perturbations, only $\Psi_{0}$ and $\Psi_{4}$ are invariant under tetrad rotations and diffeomorphisms. It is obtained that
\begin{align}
\Psi_0 &= \frac{M^2A_t^2\,(i+\cos\theta)^2}{8\,\Lambda(\theta)^3\sin^2\theta}\,\delta\, l^a l^b\nabla_a\nabla_b \chi + \mathcal{O}(\delta^2)
= \frac{M^2r^4A_t^4\,(i+\cos\theta)^2}{2\,\Lambda(\theta)^5B^4\sin^2\theta}\,\delta\, k_+^a k_+^b\nabla_a\nabla_b \chi + \mathcal{O}(\delta^2),\label{Psi0}\\
\Psi_4 &= \frac{2M^2\,(i+\cos\theta)^2}{\Lambda(\theta)\,A_t^2\sin^2\theta}\,\delta\, n^a n^b\nabla_a\nabla_b \chi + \mathcal{O}(\delta^2)
= \frac{2M^2\,(i+\cos\theta)^2}{\Lambda(\theta)\,A_t^4\sin^2\theta}\,\delta\, k_-^a k_-^b\nabla_a\nabla_b \chi + \mathcal{O}(\delta^2),\label{Psi4}
\end{align}
where in the second equalities $k_\pm:=-\partial_t\pm\frac{B}{2}\partial_r$ are the null vectors of the base $g_{ab}$ obtained by projecting $l$ and $n$, so that $l^a=\frac{2r^2A_t}{\Lambda B^2}k_+^a$ and $n^a=A_t^{-1}k_-^a$. Both scalars are proportional to $\chi$, meaning that in \eqref{Escalar eq}, $\chi$ is gauge invariant.  In both cases, we can relate to the standard separation of variables used to solve the Teukolsky master equation presented in the Appendix \ref{Apendice B}, and identify $U_l(t, r)$.

\medskip

 \subsubsection{Linearized Einstein's equations}
We solve Einstein's equations perturbatively,  expanding the Ricci tensor as
$$
R_{\mu\nu} = R_{\mu\nu}^{(0)} + \delta R_{\mu\nu}^{(1)} + \mathcal{O}(\delta^2)
$$
and solve the equations at linear order in $\delta$,  so that
$$
R_{\mu\nu}^{(1)} = 0.
$$

Inserting the metric \eqref{eq:define_pert_metric}
gives the following system of equation
\begin{align}\
\label{eq 1} & \partial_\theta(\nabla_b \mathcal{T}^b) = 0\\
\label{eq 2} & \sin^2\theta \partial_\theta(\epsilon_{ab}\mathcal{T}^b) + \cot\theta \nabla_a\chi = 0\\
\label{eq 3}& \epsilon^{ab}\nabla_a\mathcal{T}_b - \frac{\cos^2\theta}{\sin^3\theta}\partial_\theta(\frac{\chi}{\cos\theta}) = 0.
\end{align}
Here $\epsilon_{ab}$ is the antisymmetric pseudo-tensor, where  $\epsilon_{tr}=\sqrt{-\det g_{ab}}$ and all the indices are lowered and raised with the 2-dimensional metric $g_{ab}$.

Acting with $\partial_\theta$ on \eqref{eq 3} and inserting \eqref{eq 2} gives the scalar equation
\begin{equation} \label{Escalar eq}
\square_2\chi + \frac{\sin^3\theta}{\cos\theta}\partial_\theta\qty[\frac{\cos^2\theta}{\sin^3\theta}\partial_\theta(\frac{\chi}{\cos\theta})] = 0.
\end{equation}
    Here and in the following,  $\square_2$ is the nearly-$\adstwo$ Laplacian, 
\begin{equation}\label{nearlyLaplacian0}
\square_2 =\frac{4 r ^2 \mathcal{L}' }{B^{3}}\partial_{t}-\frac{4 r ^2 }{B^2}\partial^{2}_{t}+\frac{4 r ^3 }{B}\partial_{r}+r ^2 \partial^{2}_{r}.
\end{equation}

Since the scalar $\chi$ determines the Weyl scalars $\Psi_0$ and $\Psi_4$, we focus on the scalar equation~\eqref{Escalar eq} in the following. 

\subsubsection{The angular part}\label{ang_part}
The coefficients in \eqref{Escalar eq} are singular near the poles $\theta = 0, \pi$.
This forces nonsingular solutions $\chi$ to vanish suitably at these points, which guarantees that the metric~\eqref{eq:define_pert_metric} is free of conical singularities at these points. 

One way to see this is to separate variables as in
$$
\chi(x^{a},\theta) = \sin^2\theta S(\theta) \tilde\chi(x^a),
$$
which leads to the equation
\begin{equation}\label{angulareq}
\partial^{2}_{\theta}S + \cot\theta\,  \partial_{\theta}S + \left( K - \frac{4}{\sin^2\theta} \right)S = 0.
\end{equation}
which reduces to Legendre's equation with $x = \cos(\theta)$.
For the solution to be free of conical singularities, we need that $ S(\theta)$ is bounded for $\theta$ near $0, \pi$, such that $\chi(x^a,\theta)$ tends to zero as $\theta$ approaches $0,\pi$. The Legendre equation for $S$ has bounded solutions as soon as $K = l(l+1)$ for some $l\in \mathbb{N}$ with $l \ge 2$. They are of the form
\begin{equation}
S(\theta) = c_1 P_l^{(2)}(\cos\theta)
\end{equation}
with $P_l^{(m)}$ the associated Legendre function of the first kind. 

For later use, we recall that
\begin{equation}
\label{eq:associated-legendre}
P^{(2)}_l(\cos(\theta)) = \sin^2(\theta) p_l(\cos(\theta))
\end{equation}
with $p_l$ some polynomial of degree $l-2$. 
thus, the general solution has the form 
$$
\chi(x^{a},\theta) = \sin^4 (\theta) \sum_{l =2}^\infty  \chi_l (x^{a}) p_l(\cos(\theta)).
$$
where each mode $\chi_l$ satisfies 
the wave equation
\begin{equation}\label{wave-eq}
\square_2\tilde\chi(\vb*{x}) = l(l+1)\tilde\chi(\vb*{x}) 
\end{equation}
for the nearly-$\adstwo$ Laplacian \eqref{nearlyLaplacian0}.
For the cases $\mathcal{L}$ constant and equal to zero, the equation \eqref{wave-eq} becomes the radial Teukolsky equation for axisymmetric perturbations.

\subsection{Killing Vector Fields and Energy Formalism}
\label{sec:killing-fields}
In this section, we aim to obtain identities on the relation between a vector field $\zeta$ parametrized by a time-dependent function $\epsilon$ as in Section~\ref{Extensions}, and the perturbed $\adstwo$-metric that contains the free function $\cL$.  This first part is key to develop the energy formalism for the equation \eqref{Escalar eq}.

We will use a vector field $\zeta$ parametrized as in Section~\ref{Extensions} by a $t$-dependent function $\epsilon$. Recall that we took
\begin{equation}\label{general-Killin-2}
\zeta = \qty(\epsilon(t) + \frac{\epsilon''(t)}{2r^2 + \mathcal{L}(t)})\partial_t - r\epsilon'(t)\partial_r
- \frac{2r\epsilon''(t)}{2 r^2 + \mathcal{L}(t)}\partial_\phi + c_\phi\partial_\phi,
\end{equation}
where $\epsilon(t)$ is an arbitrary function of time related to the time reparametrization symmetry and $c_\phi$ is an arbitrary constant associated with the angular invariance.

\subsubsection{The deformation tensor}

 First,  we begin by computing the components of the Lie derivative along the vector field $\zeta$, 
 we obtain through direct computation that the components of the  deformation tensor $\pi$ are given by
 
 \begin{equation}\label{td1}
\pi_{rr}=\zeta^r\partial_r g_{rr}+ 2g_{rr}\partial_r\zeta^{r}=-r\epsilon'\partial_r\left(\frac{1}{r^2} \right)+\frac{2}{r^2} \partial_{r}(-r\epsilon')=0, \end{equation}
 \begin{equation}
 \pi_{tr}=g_{rr}\partial_t \zeta^r +g_{tt}\partial_r \zeta^t  =-\frac{\epsilon''}{r}-\frac{B^2}{4r^2}\partial_r\left(\epsilon +\frac{\epsilon''}{B}\right)=0,
 \end{equation}
  and
 \begin{equation}\label{td2}
 \begin{aligned}
 \pi_{tt}&=\zeta^r\partial_r g_{tt}+ \zeta^t\partial_t g_{tt}+2g_{tt}\partial_t\zeta^{t}\\
 &= -r\epsilon'\left(\frac{-32Br^3+8B^2r }{16r^4}\right)-\left(\epsilon +\frac{\epsilon''}{B} \right)\frac{B\mathcal{L}'}{2r^2}-\frac{B^2}{2r^2}\left(\epsilon'+\frac{\epsilon'''B-\epsilon''\mathcal{L}'}{B^2} \right)\\
 &= B\epsilon'\left(2-\frac{B}{r^2}\right)-\frac{B}{2r^2}\left(\epsilon \mathcal{L}'+\epsilon'''\right)=-\frac{B}{2r^2}\delta_{\epsilon}\mathcal{L}.
 \end{aligned}
 \end{equation}
Here $B$ is defined in \eqref{B-def} and, using the definition of $\delta_{\epsilon}\mathcal{L}$ in \eqref{transformation},
 \[\delta_{\epsilon}\mathcal{L}=\epsilon'''+2\mathcal{L}\epsilon'+\mathcal{L}'\epsilon.\]
The last component of the deformation tensor $\pi_{tt}$ reveals the \emph{conditional character} of $\zeta$; that is, for a given $\epsilon,$ the field $\zeta[\epsilon]$ is a Killing field of $g_{ab}$ if and only  $\delta_{\epsilon}\mathcal{L}=0$. For every profile $\mathcal{L}(t)$ this third-order linear equation has a three-dimensional space of solutions \ref{Extensions}, so Killing fields always exist.

\medskip

Using the previous results, a direct computation shows that the commutator of the wave operator $\Box_2$ with the vector field $\zeta$ can also be characterized in terms of the symmetry term $\delta_{\epsilon}\mathcal{L}$, as follows

\begin{equation*}
[\Box_2,\zeta]=\delta_{\epsilon}\mathcal{L}\left(-\frac{8r^2}{B^3}\partial_t^2+\frac{12r^2\mathcal{L}'}{B^4}\partial_t+\frac{4r^3}{B^2}\,\partial_r\right)-\frac{4r^2}{B^3}(\delta_{\epsilon}\mathcal{L})'\partial_t .
\end{equation*}

Thus, $\delta_{\epsilon}\mathcal{L}=0$ implies that 
$[\Box_2,\zeta]=0$. Moreover, since $\zeta$ is independent of $\theta$ and 

\[\rho_{\theta}[\cdot]:=\frac{\sin^3\theta}{\cos\theta}\partial_\theta\qty[\frac{\cos^2\theta}{\sin^3\theta}\partial_\theta\left(\frac{\cdot}{\cos\theta}\right)]\] 
is independent of $(r,t)$, we also have 
\[[\rho_{\theta},\zeta]=0.\]
The goal is now to develop the energy formalism for equation \eqref{Escalar eq}, and to characterize $\delta_{\epsilon}\mathcal{L}$ in such a way that allows us to establish a proper coercivity criterion. In the following section, we establish a unified identity for the energy density, depending on $\epsilon$, and derive an exact coercivity criterion for the case in which $\zeta$ is a Killing vector field.

\subsubsection{Stability from energy conservation}\label{energy_formalins}

\medskip

Recall that the full perturbation equation for the nearly-$\adstwo$ Laplacian \eqref{nearlyLaplacian0} for $\chi$ depending on $t,r, \theta$ is given by
\begin{equation}\label{fullwaveeq}
\Box_2\chi+\frac{\sin^3\theta}{\cos\theta}
\partial_\theta\left[\frac{\cos^2\theta}{\sin^3\theta}\partial_\theta\left(\frac{\chi}{\cos\theta}\right)
\right]=0, \quad (t,r,\theta) \in \mathbb{R} \times (0,\infty) \times (0,\pi).
\end{equation}
Expanding the angular part gives 
\begin{equation}\label{fullwaveeq*}
\Box_2 \chi + \chi_{\theta\theta} -3\cot \theta \chi_\theta -2\chi =\Box_2 \chi+ \frac{1}{s(\theta)}\partial_\theta(s(\theta)\chi_\theta)-2\chi=0, 
\end{equation}
with $s(\theta)=\sin^{-3}(\theta),$  and $(t,r,\theta) \in (0,\infty)^2 \times (0,\pi)$. 

\medskip

Writing its angular part in Sturm–Liouville form $\rho_{\theta}\chi= \frac{1}{s(\theta)}\partial_\theta(s(\theta)\chi_\theta)-2\chi$, we can identify the weight that makes the wave-type operator formally self-adjoint, given by
\[s(\theta)\sqrt{-\bar g}d\theta dtdr:= \sin^{-3}(\theta)\frac{B}{2r^2}d\theta dtdr.\]
Here and in the following, $\bar g$ denotes the determinant of the $t,r$-metric $g_{a b}$.
 
If $\chi$ and its derivatives vanish suitably at the poles, the equation \eqref{fullwaveeq*} matches the Euler-Lagrange equation of the Lagrangian
\begin{equation}\label{fulllagrangian}
\Lgr[\chi]
=
\frac{\sqrt{-\bar g}}{\sin^3(\theta)}\left[-\frac12
g^{ab}\partial_a\chi\partial_b\chi
-\frac12
(\chi_{\theta}^2+2\chi^2)\right],
\qquad a,b\in\{t,r\}.
\end{equation}
\begin{rem}
Equation \eqref{fullwaveeq*} is variational and is exactly  the Euler–Lagrange equation associated with the weighted functional \eqref{fulllagrangian}. Consequently, the corresponding stress-energy tensor should be interpreted as the natural energy-momentum tensor associated with the effective perturbation equation rather than with the full Einstein-Hilbert action. \end{rem}
The associated energy-momentum tensor on the $(t,r)$-sector is
\begin{equation}\label{fullstress}
T_{ab}=\partial_a\chi\,\partial_b\chi-\frac12\,g_{ab}\left(g^{cd}\partial_c\chi\,\partial_d\chi+\chi_\theta^2+2\chi^2\right),\qquad a,b\in\{t,r\},
\end{equation}
and the explicit  components are given by 

\begin{equation}\label{fullstress1}
\begin{aligned}
&~{}T_{tt}=\frac12\chi_t^2+\frac{B^2}{8}\chi_r^2+\frac{B^2}{8r^2}\left(\chi_\theta^2+2\chi^2\right),\\ 
&~{}T_{rr}=\frac{1}{2}\chi_r^2+\frac{2}{B^2}\chi_t^2-\frac{1}{2r^2}\left(\chi_\theta^2+2\chi^2\right), \\
&~{} T_{tr}=\chi_t\chi_r .
\end{aligned}
\end{equation}
 We regard \eqref{fulllagrangian} as the variational functional associated with equation \eqref{fullwaveeq*}. The resulting stress tensor should therefore be understood as an analytic energy functional for the perturbation equation.

\medskip

In order to propose the energy functional, recall the generalized Killing vector field
\begin{equation}
\zeta
=
\left(
\epsilon+\frac{\epsilon''}{2r^2+\mathcal{L}}
\right)\partial_t
-r\epsilon'\partial_r
+
\left(
-\frac{2r\epsilon''}{2r^2+\mathcal{L}}
+c_\phi
\right)\partial_\phi.
\end{equation}
Assuming axisymmetry, $\partial_\phi\chi=0,$ the $\phi$--component does not contribute dynamically. We define the component of the current as 
\[J^a=-T^a{}_b\zeta^b, \quad \mbox{and}  \qquad J^\theta:=-\chi_\theta\,\zeta(\chi),
\]
then the explicit components are given by
\begin{equation*}
\begin{aligned}
& J^t=\frac{4r^2}{B^2}(T_{tt}\zeta^t+T_{tr}\zeta^r),\\
& J^r=-r^2(T_{rt}\zeta^t+T_{rr}\zeta^r),\\
&J^{\theta}= -\chi_{\theta}\left(\frac{B\epsilon+\epsilon''}{B}\chi_t-r\epsilon'\chi_r \right),
\end{aligned}
\end{equation*}
and the  divergence satisfies 
\begin{equation}\label{eq:divergencia}
\nabla_a J^a=-(\nabla^aT_{ab})\zeta^b-\frac{1}{2}T^{ab}\pi_{ab}.
\end{equation}
For any $\chi$ we have
\[s(\theta)\nabla^{a}T_{ab}=s(\theta)(\Box_2\chi+\rho_\theta\chi)\partial_b\chi-\partial_{\theta}(s(\theta)\chi_{\theta}\partial_b \chi), \qquad b \in \{ t,r\}, \]
furthermore, since \[\nabla_aJ^a=\frac{1}{\sqrt{-\bar g}}\partial_a(\sqrt{-\bar g}J^a)\] 
hence, the equation \eqref{eq:divergencia} gives rise to the following identity 
\begin{equation}\label{eq:divergencia*}
\nabla_a(s(\theta)J^a)+\partial_\theta(s(\theta)J^\theta)=-s(\theta)(\Box_2\chi+\rho_\theta\chi)\zeta(\chi)-\frac{1}{2}s(\theta)T^{ab}\pi_{ab},
\end{equation}
using \eqref{td2}, the last term above  is given by
\begin{equation}
T^{ab}\pi_{ab}=(g^{tt})^2T_{tt}\pi_{tt}=\frac{16r^4}{B^4}T_{tt}\big(-\frac{B}{2r^2}\delta_{\epsilon}\mathcal{L}\big)= -\frac{8r^2}{B^3}\delta_{\epsilon}\mathcal{L}T_{tt}.
\end{equation}
The divergence equation \eqref{eq:divergencia*} gives us the following key quasi-continuity equation that describe the divergence identity with angular flux
 \begin{equation}\label{eq:divergencia**}
 \begin{aligned}
 & \partial_t\left(\sqrt{-\bar g}s(\theta)J^t \right)+\partial_r\left(\sqrt{-\bar g}s(\theta)J^r \right)+\partial_{\theta}\left(\sqrt{-\bar g}s(\theta)J^{\theta} \right)\\
 & \qquad =-\sqrt{-\bar g}s(\theta)(\Box_2 \chi+\rho_{\theta}\chi)\zeta(\chi)+\frac{2s(\theta)}{B^2}\delta_{\epsilon}\mathcal{L}T_{tt}.
 \end{aligned}
 \end{equation}
 Consequently, whenever $\delta_{\epsilon}\mathcal{L}=0$, $\zeta$ is a Killing vector of $g_{ab}$, $\pi_{ab}=0$, and the right-hand side of \eqref{eq:divergencia*} vanishes on solutions of \eqref{fullwaveeq}.
Now, consider the spacetime regions given by
\[ 
\mathcal{D}_{t_0}:= \{ (t,r,\theta): 0 \leq t \leq t_0, r\in (0,\infty), \theta \in (0,\pi)\},
\]
and the time slice,  \[\Sigma_{t_0} = \{ (t_0,r,\theta)\in D_{t_0}\},  \] then the future unit normal vector is $n=\frac{2r}{B}\partial_t,$  and we get in particular that 
\[
\sqrt{-\bar g}\,J^t=-\sqrt{-\bar g}\,g^{tt}T_{tb}\zeta^b=\frac{B}{2r^2}\frac{4r^2}{B^2}T_{tb}\zeta^b=\frac{2}{B}\big(T_{tt}\zeta^t+T_{tr}\zeta^r\big). 
\]
Integrating \eqref{eq:divergencia**} over $(r,\theta)\in(0,\infty)\times(0,\pi)$, the $\partial_r$ and $\partial_\theta$ terms become boundary terms that vanish under suitable assumptions, and the quantity whose time derivative remains defines the following energy formulation

\begin{equation}\label{energy-g}
\begin{aligned}
E_\epsilon(t)&~{}:=\int_{\Sigma_t}T_{ab}n^a\zeta^b\frac{dr}{r}s(\theta)d\theta\\
&~{}=\int_0^\pi\int_0^\infty \sqrt{-\bar g}s(\theta)J^tdrd\theta\\
&~{}=\int_0^\pi\int_0^\infty s(\theta)\frac{2}{B}\Big[\zeta^tT_{tt}+\zeta^rT_{tr}\Big]drd\theta
\end{aligned}
\end{equation}
 The equation \eqref{energy-g} is precisely the geometric version of an energy law for a hyperbolic equation. 
 Inserting the components of the stress tensor given in \eqref{fullstress1}  we get that the functional of energy formulation is given by 
\begin{equation}\label{energy-explicit}
E_\epsilon[\chi]=\int_0^{\pi}\int_{0}^{\infty} s(\theta)\left[\frac{B\epsilon +\epsilon''}{B^2}\chi_t^2 +\frac{B\epsilon +\epsilon''}{4}\chi_r^2+\frac{B\epsilon +\epsilon''}{4r^2}(\chi_{\theta}^2+2\chi^2)-\frac{2r\epsilon'}{B}\chi_t\chi_r\right]drd\theta.
\end{equation}
 
If we assume finite-energy boundary conditions, namely 
 \begin{equation}\label{bc}
 \begin{cases}
 &\sqrt{-\bar g}\,\sin^{-3}\theta\,J^r\to0 \quad \mbox{as} \quad r\to0^+ \wedge \, r\to\infty\\
& \sqrt{-\bar g}\sin^{-3}\theta \, J^\theta\to0 \quad \mbox{as} \quad \theta\to0,\pi, 
 \end{cases}
 \end{equation}
 then, for each choice of the vector field $\zeta$ parametrized by  $ \epsilon(t)$, from the continuity equation \eqref{eq:divergencia**} we get 
\begin{equation}\label{master_energy}
\frac{d}{dt}E_\epsilon[\chi](t)=\delta_{\epsilon}\mathcal{L}(t)\int_0^\pi\int_0^\infty s(\theta)\left[\frac{\chi_t^2}{B^2}+\frac{\chi_r^2}{4}+\frac{1}{4r^2}(\chi_{\theta}^2+2\chi^2) \right]drd\theta.
\end{equation}
Both boundary conditions are natural if $\chi$ is a perturbation of the metric. For the first one,  if $\mathcal{L}\neq 0$ the limit $r\to0^+$ is a conformal boundary and these are reflecting conditions; for $\mathcal{L}= 0$ it is the Poincaré horizon and they are decay assumptions on the initial data. As pointed out before, the second boundary condition corresponds to the absence of conical singularities.

\medskip 

Returning to \eqref{master_energy}, if we choose $\epsilon$ such that  $\delta_{\epsilon}\mathcal{L}=0$, from \eqref{master_energy} we obtain that the functional  $E_{\epsilon}$ is a conserved energy, indeed
\begin{equation}\label{conservation_new}
E_\epsilon[\chi](t)=E_\epsilon[\chi](0)\qquad\text{for all }t .
\end{equation}

 \begin{rem} 
 If $\delta_\eps\Lc=0$, then $\zeta$ commutes with $\Box_2+\rho_\theta$, so $\zeta\chi$ is again a solution and $E_\epsilon[\zeta_{\tilde\epsilon}\chi]$ is conserved for any pair of Killing parameters $\epsilon,\tilde\epsilon$. For fixed $\mathcal{L}=\{f,t\}$, the Killing parameters form the three-dimensional space $\epsilon=(a+bf+cf^2)/f'$, so this procedure yields conserved energies of higher order in derivatives (a finite family at each order), rather than new symmetries.
  \end{rem}

\subsubsection{Coercivity of the energy}
  
We now check that the conservation of $E_{\epsilon}[\chi]$ does indeed guarantee that the perturbation $\chi$ remains bounded in a suitable norm.
For this, we have to control the last term in the definition of $E_\epsilon$, which has no definite sign. To do so, we collect all the terms of \eqref{energy-explicit} containing $\chi_t$ or $\chi_r$ into a single quadratic form. Set
\begin{equation}\label{A-def}
A_\epsilon(t,r):=B\epsilon+\epsilon''=2r^2\epsilon+\mathcal{L}\epsilon+\epsilon'',\qquad \bm{v}:=\left(\frac{2\chi_t}{B},\,\chi_r\right)^{T},
\end{equation}
so that the energy \eqref{energy-explicit} takes the form
\begin{equation}\label{energy-Q}
E_\epsilon[\chi]=\int_0^{\pi}\int_{0}^{\infty} s(\theta)\left[Q_\epsilon[\chi]+\frac{A_\epsilon}{4r^2}(\chi_{\theta}^2+2\chi^2)\right]drd\theta,
\end{equation}
where
\begin{equation}\label{CF}
Q_\epsilon[\chi]:=\frac{A_\epsilon}{B^2}\chi_t^2-\frac{2r\epsilon'}{B}\chi_t\chi_r+\frac{A_\epsilon}{4}\chi_r^2=\frac14\,\bm{v}^{T}M_\epsilon\bm{v},\qquad M_\epsilon(t,r):=\begin{pmatrix} A_\epsilon & -2r\epsilon'\\ -2r\epsilon' & A_\epsilon\end{pmatrix}.
\end{equation}
The matrix $M_\epsilon$ has eigenvectors $(1,1)^T$ and $(1,-1)^T$ for every $(t,r)$, with eigenvalues
\begin{equation}\label{eigenvalues}
\lambda_\mp:=A_\epsilon\mp2r\epsilon',
\end{equation}
respectively.
Therefore $Q_\epsilon$ is positive definite at $(t,r)$ if and only if
\begin{equation}\label{lambda_min}
\lambda_{\min}(t,r):=\min\{\lambda_+,\lambda_-\}=A_\epsilon-2r|\epsilon'|>0,
\end{equation}
or equivalently, in terms of the trace and the determinant of $M_\epsilon$,
\begin{equation}\label{condition*}
A_\epsilon>0\qquad\mbox{and}\qquad A_\epsilon^2-4r^2(\epsilon')^2>0.
\end{equation}
In the orthonormal frame $n=\frac{2r}{B}\partial_t$, $e=r\partial_r$, the $(t,r)$ part of $\zeta$ reads
$\zeta=\frac{A_\epsilon}{2r}\,n-\epsilon'\,e$, so that
\[
g(\zeta,n)=-\frac{A_\epsilon}{2r},\qquad
g(\zeta,\zeta)=-\frac{A_\epsilon^2-4r^2(\epsilon')^2}{4r^2}=-\frac{\lambda_+\lambda_-}{4r^2}.
\]
Hence \eqref{condition*} are precisely the conditions that $\zeta$ be timelike and future directed with respect to $n$. Moreover, for any sign of $\lambda_{\min}$,
\begin{equation}\label{quadraticform}
Q_\epsilon[\chi]\geq\frac14\lambda_{\min}|\bm{v}|^2=\lambda_{\min}\left(\frac{\chi_t^2}{B^2}+\frac{\chi_r^2}{4}\right).
\end{equation}

\medskip

The dependence on $r$ is made explicit by completing the square in \eqref{eigenvalues}:
\begin{equation}\label{lambda-square}
2\epsilon\,\lambda_\pm=(2r\epsilon\pm\epsilon')^2+C_\epsilon,\qquad C_{\epsilon}:= 2\epsilon \epsilon''-(\epsilon')^2+2\mathcal{L}\epsilon^2.
\end{equation}
Since $\lambda_{\min}\sim2r^2\epsilon$ as $r\to\infty$, the condition $\epsilon\geq0$ is necessary. For $\epsilon>0$, \eqref{lambda-square} gives $2\epsilon\lambda_{\min}=(2r\epsilon-|\epsilon'|)^2+C_\epsilon$, whose minimum over $r>0$ is attained at $r_*:=|\epsilon'|/(2\epsilon)$. Thus, on a slice where $\epsilon>0$:
\begin{itemize}
\item If $C_\epsilon>0$, then $\lambda_{\min}\geq C_\epsilon/(2\epsilon)>0$, and $Q_\epsilon$ is positive definite for all $r>0$.
\item If $C_\epsilon=0$ and $\epsilon'\neq0$, then $Q_\epsilon$ is non-negative but degenerates at $r=r_*(t)$. If $C_\epsilon=0$ and $\epsilon'=0$, then $\lambda_{\min}=2r^2\epsilon$, which degenerates only as $r\to0$.
\item If $C_\epsilon<0$, then $\lambda_{\min}<0$ at $r=r_*$ when $\epsilon'\neq0$, and for small $r$ when $\epsilon'=0$. There $\zeta$ is not future timelike and $Q_\epsilon$ takes negative values.
\end{itemize}
In the case $\epsilon>0$, $C_\epsilon\geq0$, $\epsilon'^2+C_\epsilon>0$, we also compare $Q_\epsilon$ with its diagonal part $\frac14A_\epsilon|\bm{v}|^2$. Here $A_\epsilon>0$, and the ratio $2r|\epsilon'|/A_\epsilon$ is maximal at $r_M=\sqrt{(\mathcal{L}\epsilon+\epsilon'')/2\epsilon}$, which gives
\begin{equation}\label{C0-def}
C_0(t):=\inf_{r>0}\frac{\lambda_{\min}(t,r)}{A_\epsilon(t,r)}=1-\frac{|\epsilon'|}{\sqrt{2\epsilon(\mathcal{L}\epsilon+\epsilon'')}}=1-\frac{|\epsilon'|}{\sqrt{\epsilon'^2+C_{\epsilon}}}\in[0,1].
\end{equation}
Thus $C_0(t)>0$ whenever $C_\epsilon>0$, and $C_0(t)=0$ when $C_\epsilon=0$ and $\epsilon'\neq0$, in which case $r_M=r_*$.
\begin{rem}
Note that $C'_{\epsilon}=2\epsilon\,\delta_{\epsilon}\mathcal{L}$, so $C_\epsilon$ is constant for Killing parameters; the bounds below are pointwise in $t$ and do not require $\delta_\epsilon\mathcal L=0$.
\end{rem}
Since the angular part in \eqref{energy-Q} is non-negative whenever $A_\epsilon\geq0$, using \eqref{quadraticform} in \eqref{energy-Q} we obtain the pointwise coercivity
\begin{equation}\label{coercividad}
E_\epsilon[\chi](t) \geq \int_0^\pi \int_0^\infty s(\theta)\left[\frac14\lambda_{\min}|\bm{v}|^2+\frac{A_\epsilon}{4r^2}\left(\chi_\theta^2+2\chi^2\right)\right]drd\theta.
\end{equation}

Using the uniform bound $C_0(t)$ in $r$, we obtain the criterion
\begin{equation}\label{coercividad*}
E_\epsilon[\chi](t) \geq C_0(t) \int_0^\pi \int_0^\infty s(\theta)\,A_\epsilon\left[\frac14|\bm{v}|^2+\frac{1}{4r^2}\left(\chi_\theta^2+2\chi^2\right)\right]drd\theta .
\end{equation}

It is worth emphasizing that, given the general framework in which the energy formalism has been presented up to this point, estimates \eqref{coercividad} and \eqref{coercividad*} will provide useful criteria across various problem settings. In particular, assuming $\delta_{\epsilon}\mathcal{L}=0$ and $\epsilon>0$. If $C_\epsilon>0$, the perturbation equation admits a conserved energy which is coercive in the sense of \eqref{coercividad*}, with $C_0(t)>0$. If $C_\epsilon=0$, the energy is conserved and non-negative but degenerate on the curve $r=r_M(t)$; this is the case of $\epsilon=1$ for $\mathcal{L}=0$, where $r_M=0$ is the horizon, which suggests that horizon instabilities of Aretakis type may still be present. If $C_\epsilon<0$, the energy is conserved but not sign-definite: for instance, for $\mathcal{L}=-\kappa^2/2$ and $\epsilon=1$ one has $C_1=-\kappa^2$ and $B$ changes sign at $r=\kappa/2$. Positivity of $\epsilon$ alone is therefore not sufficient. The conservation law is tied to the Killing symmetries of the Schwarzian background, and through \eqref{master_energy} the same construction quantifies the failure of conservation when $\delta_\epsilon\mathcal{L}\neq0$. The aim of the following sections is precisely to identify what characterizations we can obtain from this formalism, depending on the background under consideration, starting with the near-NHEK case, but also assuming more general settings for the background $\mathcal{L}$.

\subsection{Applications to specific backgrounds}
\label{sec:vector_fields_aplications} 
 
We now use \eqref{master_energy} and its consequences on several classes of backgrounds: the near-NHEK solution, the case of general $\cL$, and the KdV soliton-solution.  
 
 \subsubsection{The Thermal Case: near-NHEK}\label{sec:thermal}
 Taking into account that the energy defined in the previous section is, in a suitable sense, a general density in terms of the fields $\mathcal{L}(t), \epsilon(t)$, let us consider the thermal background   $\mathcal{L}(t)=-\kappa^2/2,$ with   $\epsilon =\cosh(\kappa t)$ is one of the Killing parameters.  In this setting, it follows that $\delta_{\epsilon}\mathcal{L}=0,$ implying that $\zeta$ is indeed a Killing vector field for the nearly-$\adstwo$ Laplacian, therefore by \eqref{master_energy} the energy $E_{\epsilon}[\chi](t)$ defined in \eqref{energy-explicit} is conserved. Furthermore, a direct computation yields the quantities appearing in the coercivity estimate \eqref{coercividad}, as shown below
 
 \begin{equation}
 \begin{aligned}
 &~{}B= 2r^2-\kappa^2/2 , &  A_\epsilon= 2\cosh(\kappa t)(r^2+\kappa^2/4)>0,\\
   &~{}  B^2= 4(r-\kappa/2)^2(r+\kappa/2)^2, 
 \end{aligned}
 \end{equation} 
 meanwhile
 \begin{equation}
 \lambda_{\min}=2\cosh(\kappa t)(r^2+\kappa^2/4)-2r\kappa|\sinh(\kappa t)| \geq 2\cosh(\kappa t)(r-\kappa/2)^2
 \end{equation}
 Simplifying, we obtain
 
 \begin{equation*}
 \begin{aligned}
& \frac{\lambda_{\min}}{B^2}\geq \frac{\cosh(\kappa t)}{2(r+\kappa/2)^2}, \qquad \frac{\lambda_{\min}}{4}\geq \frac{1}{2}\cosh(\kappa t)(r-\kappa/2)^2,\\
& \frac{A_\epsilon}{4r^2}= \cosh(\kappa t )\left(\frac{r^2+\kappa^2/4}{2r^2}\right),
 \end{aligned}
 \end{equation*}
 therefore using \eqref{coercividad}, for finite-energy data on the whole slice, 
 
 \begin{equation}\label{coercive-norm}
E_\epsilon[\chi](t)\geq\cosh(\kappa t)\|\chi(t)\|^2,
\end{equation}
where the norm in this case is define by
\begin{equation}
\|\chi\|^2:=\int_0^\pi\int_0^\infty\sin^{-3}\theta\left[\frac{1}{2(r+\kappa/2)^2}\chi_t^2+\frac{(r-\kappa/2)^2}{2}\chi_r^2+\frac{r^2+\kappa^2/4}{2r^2}(\chi_\theta^2+2\chi^2)\right]drd\theta .
\end{equation}
Consequently, since $\delta\mathcal{L}=0$, by \eqref{conservation_new} and \eqref{coercive-norm},
\[
\|\chi(t)\|^2\leq \frac{E_{\epsilon}[\chi](0)}{\cosh(\kappa t )}
\]
Thus, in the thermal background, the norm $\norm{\chi}$ decays exponentially in $t,$ at least at the rate $\kappa$ set by the temperature. Only the weight of $\chi_r^2$ degenerates at $r=\frac{\kappa}{2}$; in particular, the weighted norms of $\chi_t$ and of the angular terms decay at this rate also at $r=\frac{\kappa}{2}$, and the full local energy decays at the same
rate on any region $\{|r-\kappa/2| \geq\delta, \, r \leq R \}$.
\medskip

\begin{rem}
The estimate above uses the point-wise bound on $\lambda_{\min}$, and the positivity of $A_\epsilon$ is checked directly; the constant $C_\epsilon$ plays no role in it. It enters only if one uses instead the infimum in $r$, 
\[C_0(t)= 1-\frac{|\kappa||\sinh(\kappa t)|}{\sqrt{\kappa^2\sinh^2(\kappa t)+\kappa^2}}, \qquad  C_{\epsilon}= \kappa^2,\] 
in \eqref{coercividad*}. This gives weights that do not degenerate at $r=\kappa/2$, but since $C_0(t)\cosh(\kappa t)=e^{-\kappa|t|}$ the resulting bound,
\[
\int_0^\pi\int_0^\infty\sin^{-3}\theta(r^2+\kappa^2/4)\left[\frac{2\chi_t^2}{B^2}+\frac{\chi_r^2}{2}\right]drd\theta \leq \ e^{\kappa|t|}E_\epsilon[\chi](0),
\]
allows growth of that stronger norm. The point-wise bound is therefore the one that yields stability and decay, while the infimum is the one to use when non-degenerate control at $r=\kappa/2$ is needed. The infimum is attained precisely at $r=\kappa/2$.
\end{rem}

\begin{rem} For every $\epsilon$ and every background $\mathcal{L}(t$) of the
family, the energy $E_{\epsilon}$ of the field $\zeta[\epsilon]$ satisfies the identity \eqref{master_energy}, so that three criteria can be
separated. Conservation holds if and only if $\delta_{\epsilon}\mathcal{L}=0$, that is, if $\zeta[\epsilon]$  is a Killing field of the background.
Coercivity is a pointwise property in time, controlled by $\lambda_{\min}$; it holds on the whole slice when $\epsilon >0 $ and $C_{\epsilon} >0$. Decay, finally, is not produced by
conservation alone; it appears when a conserved and coercive energy carries a Killing parameter that grows in time, as in the thermal case, where the factor $\cosh(\kappa t)$ turns conservation into exponential decay of the
norm $\norm{\chi}$. The same identity with $\epsilon=1$ and $\delta_{\epsilon}\mathcal{L}=\mathcal{L}'$ gives the Lyapunov functional as will see in the  Section \ref{time-case}, and
with $\mathcal{L}=0$, $\epsilon=1$ a conserved energy.
\end{rem}

\subsubsection{Time-dependent Schwarzian background}\label{time-case} 
We now consider \eqref{Escalar eq} on a general background, with $\mathcal L=\mathcal L(t)$ a suitable function and $B=2r^2+\mathcal L$. Since the equation is not time invariant, we do not expect to find a conserved energy in general. To simplify the expressions, set
\begin{equation}\label{D-def}
D[\chi]:=\frac{\chi_t^2}{B^2}+\frac{\chi_r^2}{4}+\frac{1}{4r^2}(\chi_{\theta}^2+2\chi^2),\qquad E_\text{mod}(t):=\int_0^\pi\int_0^\infty s(\theta)\,D[\chi]\,drd\theta,
\end{equation}
so that \eqref{energy-explicit} and \eqref{master_energy} read
\[
E_\epsilon[\chi]=\int_0^\pi\int_0^\infty s(\theta)\left[A_\epsilon D[\chi]-\frac{2r\epsilon'}{B}\chi_t\chi_r\right]drd\theta,\qquad \frac{d}{dt}E_\epsilon[\chi]=\delta_\epsilon\mathcal{L}\,E_\text{mod}(t).
\]
We now take the simplest choice $\epsilon\equiv1$, for which $\zeta=\partial_t$ (up to the term in $\partial_\phi$, which does not contribute), $A_1=B$, $\delta_1\mathcal L=\mathcal L'$, and the cross term vanishes. The energy $E:=E_1[\chi]$ of this section is therefore
\begin{equation}\label{E_new}
E(t)=\int_0^\pi\int_0^\infty s(\theta)\,B\,D[\chi]\,drd\theta,
\end{equation}
and it satisfies
\begin{equation}\label{dE_new}
\frac{d}{dt}E = \mathcal L'\, E_\text{mod}(t).
\end{equation}
Precisely, this last identity reveals that $E$ is a Lyapunov functional. If $\mathcal L\geq0$, then $B>0$ for all $r>0$ and $E\geq0$; this agrees with Section \ref{energy_formalins}, since $\lambda_\pm=B$ and $C_1=2\mathcal L$. For $\mathcal{L}<0$, $E$ is not sign-definite, see for instance Section \ref{sec:thermal}.

We define $\mathcal E$ as the energy space of functions where $E$ in \eqref{E_new} is well-defined and $E$ satisfies \eqref{dE_new}. By \eqref{bc} with $\zeta=\partial_t$, this space requires
\[
\begin{aligned}
& E \hbox{ well-defined in $(t,r,\theta)\in (0,\infty)\times (0,\infty) \times (0,\pi)$}, \\
& s(\theta)\, B\, \partial_t\chi\, \partial_r \chi  \hbox{ vanishing at $r=0^+$ and as $r\to +\infty$, with $\theta,t$ free parameters},\\
&  \frac{B}{2r^2}\,s(\theta)\,\partial_t\chi\,  \partial_\theta \chi  \hbox{ vanishing at $\theta=0$ and $\theta =\pi$, with $r,t$ free parameters.}
\end{aligned}
\]
The previous conditions impose decay properties on $\chi$ and its derivatives at the boundaries. Precisely, these conditions are set to have a well-defined energy. Weaker conditions can be assumed, however the existence of a well-defined energy is not clear to us. We shall assume these finite energy conditions from now on.

There are several interesting consequences obtained directly from \eqref{dE_new}. We shall always assume that $\mathcal L(t) \geq 0$ for $t\geq 0$. 

Notice first that we recover the case $\mathcal L\equiv 0$, where $E$ is constant in time. Moreover, the sign of the derivative of $E$ only depends on the sign of $\cL'$. Therefore, in the case $\mathcal L$ decreasing in time, we get a decreasing energy $E$.

\subsubsection{Profiles with a positive lower bound}

 Now we assume that $\mathcal L$ increases towards a nonzero positive value $\mathcal L_0$. First of all, \eqref{dE_new} reads
\[
\frac{d}{dt}E =  |\mathcal L'(t)|\, E_\text{mod}(t).
\]
Since $\mathcal L$ is increasing, we have $B \geq \mathcal L(0)$. Therefore, from \eqref{dE_new},
\[
 \frac{d}{dt}E = |\mathcal L'| \int_0^\pi\int_0^\infty \frac1{B}\, s(\theta) B D[\chi]\,drd\theta  \leq  \frac{|\mathcal L' (t)|}{\mathcal L(0)} E,
\]
leading to 
$$E(t) \leq e^{(\mathcal L(0))^{-1} \int_0^t |\mathcal L' |(s)ds } E_0.$$ 
In the case where one has $\lim_{t\to +\infty}\int_0^t |\mathcal L' |(s)ds  <\infty$, one has that $E$ stays bounded for all times.

In the general case where $\mathcal L'(t)$ has any sign, but $\mathcal L(t) \geq \mathcal L_0>0$, from \eqref{dE_new} one has
\[
 \frac{d}{dt}E  \leq  \frac{ \left| \mathcal L' \right|}{\cL_0} E.
\]
Consequently, if $\int_0^\infty  |\mathcal L'(t)| dt <C_\cL<+\infty,$ we will have the stability bound $E(t) \leq E_0 e^{  C_\cL/ \cL_0}$.

\subsubsection{The KdV case}

We continue describing the perturbed NHEK case, but this time we assume  that $\mathcal L$ satisfies the KdV equation \eqref{KdV}.

First of all, notice that from \eqref{KdV} $\mathcal L(t)$ satisfies
\begin{equation}\label{kdv*}
\frac{3}{2}\mathcal{L}^2 +\mathcal{L}''+\bar{\epsilon}_{1}\mathcal{L}=0,  
\end{equation}
where we have assumed that $\mathcal L(+\infty)=0$, as well as its derivatives. 
{Regular solutions decaying at infinity require $\bar{\epsilon}_{1}:= -\omega^2<0$}, then one has soliton solutions of the form
\begin{equation}\label{soliton-solution}
\mathcal L (t) =  \frac{\omega^2 }{\cosh^2 \left( \omega \frac{(t- t_0)}{2}\right)},
\end{equation}
where $t_0$ is any free parameter. With no loss of generality, we assume $t_0=0$.  Therefore, from \eqref{dE_new} one has that $E$ is a decreasing quantity and for all $t\geq 0$,
\[
 E(t) \leq E_0.
\]
Notice that in this case $B$ vanishes as $r\to 0$ and $t\to \pm\infty$, so the coefficients in the energy are singular. Moreover, 
\begin{equation}\label{dE_new_KdV}
\frac{d}{dt}E =  \frac{-\omega^3 \sinh\left( \omega \frac{t}{2}\right)}{\cosh^3 \left( \omega \frac{t}{2}\right)}\, E_\text{mod}(t).
\end{equation}
From \eqref{dE_new_KdV}, we are in a case where $\mathcal L'<0$ and 
\[
\int_0^\infty|\Lc'|E_{\rm mod}\le E_0. 
\]
However, since the weight induced by the soliton is very strong, only a particular portion of the energy density is converging to zero. Indeed, 
\[
 \frac{d}{dt}E  = \frac{-\omega \sinh\left( \omega \frac{t}{2}\right)}{\cosh \left( \omega \frac{t}{2}\right)} \int  \frac{\mathcal L}{B}\, e(t) ,
\]
where $e:=s(\theta)BD[\chi]$ is the energy density.  Therefore, proceeding as in the previous computations, since the energy is bounded and $\frac{-\omega \sinh\left( \omega \frac{t}{2}\right)}{\cosh \left( \omega \frac{t}{2}\right)}$ does not integrate in time, we can ensure that there is a sequence $t_n\to +\infty$ such that the mild decay property $\lim_{n\to+\infty} \int  \frac{\mathcal L(t_n)}{B(t_n,r)} e(t_n) =0$ holds.

\subsubsection{A conserved energy on the soliton background}
 The previous energy $E$ defined  in \eqref{E_new} corresponds to the energy $E_{1}$ in \eqref{energy-explicit} with $\epsilon=1,$ for which $\delta \mathcal{L}=\mathcal{L}' \neq 0.$ The energy formalism presented in Section \ref{energy_formalins} allows another choice.  Using the same background 
 \[ \mathcal L (t) =  \frac{\omega^2 }{\cosh^2 \left(  \frac{\omega t}{2}\right)}, \]
 we have that $\epsilon= \omega^2-\mathcal{L}$, which is local in $\mathcal {L}$, yields a Killing vector for the equation \eqref{fullwaveeq*}, precisely because $\mathcal{L}$ solves the KdV equation with $\bar{\epsilon}=-\omega^2$. Therefore, the energy $E_{\epsilon}$ associated with this parameter is conserved. Note that for this setting, the terms appearing in the quasi-coercivity estimate \eqref{coercividad}have a definite sign; indeed, since $\mathcal{L}$ satisfies \eqref{kdv*}, then
 
\begin{equation*}
\begin{aligned}
&\mathcal{L}\epsilon-\mathcal{L}''= \mathcal{L}(\omega^2-\mathcal{L})-(\omega^2\mathcal{L}-\frac{3}{2}\mathcal{L}^2)= \frac{1}{2}\mathcal{L}^2, \\
&\epsilon=\omega^2\left(1-\sech^2\left(\frac{\omega t}{2}\right)\right)= \omega^2\tanh^2\left(\frac{\omega t}{2}\right),
\end{aligned}
\end{equation*}
therefore, the principal term $A_\epsilon$ is given by 
 \[ A_\epsilon= 2r^2\epsilon+\mathcal{L}\epsilon-\mathcal{L}''= 2r^2\omega^2\tanh^2\left(\frac{\omega t}{2}\right)+\frac{1}{2}\mathcal{L}^2 >0,\]
 meanwhile $C_\epsilon=0$, and for $\lambda_{\min}$ we get
  \begin{align}
 \lambda_{\min}= 2r^2\omega^2\tanh^2(\omega t/2)+\frac{1}{2}\mathcal{L}^2-2r|\mathcal{L}'|= \frac{1}{2}\left(2r\omega |\tanh(\omega t/2)|-\mathcal{L}\right)^2\geq 0,
 \end{align}
 so that, by \eqref{coercividad} the energy $E_{\epsilon}$ described in \eqref{energy-explicit} is nonnegative; here $C_0(t)=0$ for $t\neq0,$ thus, the energy becomes degenerate when $r = \frac{\omega}{2\cosh(\omega t/2)|\sinh(\omega t/2)|},$ and  $r \to 0$ when $t\to \pm \infty$. At the instant $t = 0,$ we have $\epsilon= \epsilon'=0$ and $A_\epsilon=\frac{\omega^4}{2}$, hence by \eqref{conservation_new}, for all $t$
 
 \begin{equation*}
 E_{\epsilon}[\chi](t)= E_{\epsilon}[\chi](0)=  \frac{\omega^4}{2} E_\text{mod}(0), 
 \end{equation*}
 where $E_\text{mod}$ is the functional defined in \eqref{D-def}.
 
\subsection{Virial methods for the NHEK case \texorpdfstring{$\cL = 0$}{L = 0}} \label{sec:virial}
In this case, from \eqref{master_energy} with $\epsilon=1$ and $\delta_\epsilon\cL = 0$, taking the expressions from Subsection~\ref{sec:thermal} with $\kappa= 0$, so that $A_1=B=2r^2$, the conserved energy reads
\begin{equation}\label{E1-NHEK}
E_1[\chi]=\int_0^\pi\int_0^\infty s(\theta)\,2r^2 D[\chi]\, drd\theta=\int_0^\pi\int_0^\infty s(\theta)\left[\frac{\chi_t^2}{2r^2}+\frac{r^2\chi_r^2}{2}+\frac12\left(\chi_\theta^2+2\chi^2\right)\right] drd\theta.
\end{equation}

To simplify the radial part of the wave equation, define, as usual $\psi:= r\chi$, $r>0$. Then $\psi$ satisfies
\begin{equation}\label{eqn}
\frac1{r^{4}} \psi_{tt}  =  \psi_{rr}   +  \frac1{r^2}\left( \psi_{\theta\theta} -3 \cot \theta \psi_\theta -2\psi\right).
\end{equation}
Since $\chi=\psi/r$, one has $\chi_t=\psi_t/r$, $\chi_\theta=\psi_\theta/r$ and $r^2\chi_r^2=\psi_r^2-\partial_r(\psi^2/r)$, so that \eqref{E1-NHEK} becomes
\begin{equation}\label{E1-psi}
\tilde E_1[\psi]:= E_1[r^{-1}\psi]=\frac12\int_0^\pi\int_0^\infty s(\theta)\left(\frac{\psi_t^2}{r^4}+\frac{2\psi^2}{r^2}+\psi_r^2+\frac{\psi_\theta^2}{r^2}\right)drd\theta
\end{equation}
This conserved energy gives a form of orbital stability. In order to get decay, we need to obtain a virial functional: a quadratic form that is finite in the energy space and whose \emph{time derivative} controls a localized norm of the perturbation.
To this end, using the strategy from \cite{MT}.

In the following computations, we shall always assume that the boundary terms are suitably made zero by finite energy assumptions. Following the lines in \cite{MT}, we need the continuity equation for the energy density in \eqref{E1-psi}: for $r>0$, $\theta\in (0,\pi)$, 
\begin{equation}\label{cons_energy}
\partial_t \left( \frac{s(\theta)}{2} \left( \frac{\psi_t^2}{r^4} +\frac{2\psi^2}{r^2} +\psi_r^2 +\frac{\psi_\theta^2}{r^2}\right)\right)= \partial_r \left(s(\theta) \psi_{r} \psi_t \right) +\partial_\theta \left(  \frac{s(\theta)}{r^2} \psi_{\theta}\psi_t \right);
\end{equation}
 we obtain the \emph{distorted conservation law} 
\begin{equation}\label{cons_mom}
\begin{aligned}
\partial_t \left( \frac{s(\theta)}{r^2} \psi_r \psi_t \right)
={}& \partial_r \left( s(\theta) \frac{\psi_t^2}{2r^2} \right)
   + s(\theta) \frac{\psi_t^2}{r^3}
   + \frac12 s(\theta)\, \partial_r (r^2\psi_r^2)
   - s(\theta)\, r \psi_r^2 \\
&+ \partial_\theta\left(s(\theta)\, \psi_\theta \psi_r \right)
   - \partial_r \left( s(\theta) \frac{\psi_\theta^2}{2} \right)
   - \partial_r\left( s(\theta)\, \psi^2 \right) .
\end{aligned}
\end{equation}
Additionally, we shall need the identity
\begin{equation}\label{cons_nada}
\begin{aligned}
\partial_t \left( \frac{s(\theta)}{r^3}  \psi \psi_t \right) = &~{}s(\theta) \frac{\psi_t^2}{r^3} +s(\theta) \left( \partial_r(r\psi\psi_r) -\psi\psi_r -r\psi_r^2  \right) \\
&~{} + \partial_\theta \left( \frac{s(\theta)}{r} \psi\psi_\theta \right) - \frac{s(\theta)}{r} \psi_\theta^2 - - \frac{2 s(\theta)}{r} \psi^2. 
\end{aligned} 
\end{equation}
The previous identities are motivated by the natural d'Alambertian operator, that even under a twisted geometry ``preserves'' a suitable energy and momentum density.

As before, we consider the energy space $\mathcal E$ as precisely the set of data $(\psi,\partial_t \psi)|_{t=0}$ and evolutions $(\psi,\partial_t \psi)(t)$, $t>0$ under which both $\psi_{r} \psi_t \sin^{-3}\theta $ and $\frac1{r^2}\sin^{-3}\theta \psi_{\theta}\psi_t$ vanish at the boundary, and $\widetilde E_1$ is well-defined. We shall prove in Appendix~\ref{app:energy_estimates} that one has the integrated local energy bound
\begin{equation}\label{virial}
 \int_0^t \int_{r,\theta}  s(\theta) \left( {\frac{1}{r}\left(\psi_\theta^2 + \psi^2\right)}\right)+  \int_0^t \int_{r,\theta}  \frac{s(\theta)}{1+r^2} \left( r^2 \psi_r^2 +\frac{\psi_t^2}{r^2} \right)  \leq C(E),
\end{equation}
revealing mild decay of finite energy solutions. Finally, one has the asymptotic stability property for finite energy solutions:
\begin{equation}\label{virial_final}
\lim_{t\to +\infty} \int_{r,\theta}\frac{r^2 s(\theta)}{1+r^2}   \left( \frac{\psi_t^2}{r^4} +\frac{\psi^2}{r^2} +\psi_r^2 +\frac{\psi_\theta^2}{r^2}\right)=0.
\end{equation}
Therefore, the solutions converge to zero in a weighted norm (note that \eqref{virial_final} differs from the norm in \eqref{E1-psi} by the additional factor $r^2/(1+r^2)$) .
This weight vanishes as $r$ approaches the horizon at zero, so it only gives information about the decay away from this boundary. 
Finally, we stress that these computations are valid when all boundary terms vanish, so we need a suitable decay of $\psi$ near $r= 0$ and as $r \to \infty$.

\subsection{Averaged and Aretakis instability in the Near-NHEK} 
\label{sec:aretakis}

Now we analyze averaged and Aretakis horizon instabilities \cite{Aretakis} for the near-NHEK family of metrics \eqref{4near-NHEK}. We shall provide an analysis in terms of growth of local energy type norms based on coordinates adapted to the event horizon. These local energy norms are less demanding than the ones introduced in previous sections, precisely to be able to capture a more detailed behavior at the near horizon itself. For the moment we shall assume $\mathcal{L}(t) = -\tfrac{\kappa^{2}}{2}$. We define
$\rho = r - \frac{\mathcal{L}}{2r} = r + \frac{\kappa^{2}}{4r}$, which maps
$r \in (\kappa/2, \infty)$ onto $\rho \in (\kappa, \infty)$, and set $\bar{r} = \rho - \kappa$,
so that $\bar{r} \geq 0$, with the event horizon located at $\bar{r} = 0$.
We shall now use ingoing Eddington--Finkelstein coordinates. Define the tortoise
coordinate $\frac{dr_*}{d\bar{r}} = \left(\bar{r}(\bar{r}+2\kappa)\right)^{-1}$, so that
$r_*(\bar{r}) = \frac{1}{2\kappa} \log \left| \frac{\bar{r}}{\bar{r}+2\kappa} \right|$. Denote
\[
v = t + r_*(\bar{r}), \qquad
\tilde{\phi} = \phi - \frac{1}{2} \log \left| \bar{r}(\bar{r}+2\kappa) \right| .
\]
The metric \eqref{4dmetric} in this case becomes
\[
ds^2 =  \Lambda(\theta) \left( -r(r+2\kappa)dv^2 +2dvdr +d\theta^2\right) + \Gamma(\theta) (d\tilde\phi +(r+\kappa)dv)^2\,.
\]
In what follows we drop the bar and write $r$ for $\bar r$, so that $r\in[0,\infty)$ and the event horizon is located at $r=0$.
 Then \eqref{fullwaveeq*} becomes
\begin{equation}\label{Escalar eq 3}
\Box_2 \chi +\frac{1}{s(\theta)}\partial_{\theta}(s(\theta) \chi_{\theta})-2\chi=0,
\end{equation}
where the nearly-$\adstwo$ Laplacian is given now by
\[
\partial_r \left(  (r^2 +2\kappa r) \partial_r  + 2\partial_v \right).
\]

\subsection{Averaged instability}

As previously mentioned, we shall work in a space slightly larger than the energy space determined in \eqref{E_new}. This will be expressed in terms of the weighted norms used to show growth of local norms. First, assume $\kappa>0$. Since $r^2+2\kappa r$ vanishes at $r=0,$ integrating \eqref{Escalar eq 3} in  $r$ from the horizon gives
\begin{equation}\label{eq:Aretakis_1}
(r^2+2\kappa r)\partial_r\chi+2\partial_v\chi=2\partial_{v}\chi|_{r=0}-\partial_r^{-1} \left( \frac{1}{s(\theta)}\partial_{\theta}(s(\theta) \chi_{\theta})-2\chi \right), \quad \partial_r^{-1}f:=\int_{0}^rfdr'. 
\end{equation}
The first term on the right-hand side is the value of $2\partial_{v}\chi$ on the horizon; it does not vanish in general. Integrating on $(r,\theta)\in[0,\infty)\times (0,\pi)$ and using \eqref{eq:Aretakis_1}
\begin{equation}\label{aretakis2}
\begin{aligned}
\frac{d}{dv} \frac12\int \frac1{r^{\frac12}} s(\theta) \chi^2  = &~{} -\frac12\int \frac1{r^{\frac12}} \chi \partial_r^{-1}\left(\partial_{\theta}(s(\theta) \chi_{\theta})-2s(\theta) \chi \right)\\
&~{} + \frac14 \int s(\theta) \partial_r(r^{\frac32}+2\kappa r^{\frac12}) \chi^2 +\int\frac1{r^{\frac12}} s(\theta) \chi \partial_{v}\chi|_{r=0}.
\end{aligned}
\end{equation}
Notice that the weight $r^{-\frac12}$ represents a concentrated operator that localizes measurements around $r=0$. 

\medskip

As a first approach, let us assume that the  term $\partial_{v}\chi|_{r=0}$ vanishes, then, integrating by parts and simplifying we get,
\[
\begin{aligned}
\frac{d}{dv} \frac12\int \frac1{r^{\frac12}} s(\theta) \chi^2  = &~{} \frac12\int \frac1{r^{\frac12}} s(\theta)\cos^2\theta \partial_\theta \left(\frac{\chi}{\cos\theta} \right) \partial_r^{-1}\left( \partial_\theta \left( \frac{\chi}{\cos\theta}\right)\right) \\
&~{} + \frac18 \int s(\theta) \left(3 r^{\frac12} + \frac{2\kappa }{r^{\frac12}} \right) \chi^2\\
\geq  &~{} \frac18\int \frac1{r^\frac32}  s(\theta)\cos^2\theta\left( \partial_r^{-1} \partial_\theta \left( \frac{\chi}{\cos\theta}\right)\right)^2 + \frac14 \kappa \int \frac1{r^{\frac12}} s(\theta)  \chi^2\\
\geq &~{}  \frac14 \kappa \int \frac1{r^{\frac12}} s(\theta)  \chi^2 .
\end{aligned}
\]
We have used here that the boundary terms $ s(\theta)(r^{\frac32}+2\kappa r^{\frac12}) \chi^2|_{r=0} $ and very importantly the term $\frac1{r^{\frac12}}\chi \partial_r^{-1}\left(\frac{1}{\cos\theta}\left( s(\theta)\cos^2\theta \partial_\theta \left( \frac{\chi}{\cos\theta}\right)\right) \right)|_{r=0}$ are zero. This is ensured if $\chi$ has sufficient decay to zero at $r=0$. Therefore,
\[
\int \frac1{r^{\frac12}} s(\theta) \chi^2 (v) \geq e^{\frac12 \kappa v} \int \frac1{r^{\frac12}} s(\theta) \chi^2 (v=0).
\]
We can conclude that there is exponential instability and growth of this energy weighted norm as $v$ tends to infinity. We emphasize that the allowed data is important. Indeed, the previous assumption is not satisfied by \emph{finite energy solutions} in the sense presented above. For instance, using the modes of Section \ref{ang_part}
\[
\chi=\sin^4\theta p_l(\cos\theta)  e^{-(l+1)\kappa v}(r+2\kappa)^{-(l+1)},
\]
solves \eqref{Escalar eq 3}, is regular at $r=0$ and decays as $r\to \infty,$ while $\int r^{-1/2}s(\theta) \chi^2$ decays like $e^{-2(l+1)\kappa v};$  This implies that, for this solution, the last term in \eqref{aretakis2} is negative and dominates, in concordance with the absence of a classical Aretakis instability in the case $\kappa>0$.

\medskip

Without any assumption on the term  $\partial_{v}\chi|_{r=0}$, we recover \eqref{aretakis2}, again, integrating by parts and simplifying we get,

\begin{equation}\label{aretakis3}
\begin{aligned}
\frac{d}{dv} \frac12\int \frac1{r^{\frac12}} s(\theta) \chi^2  = &~{} \frac18\int\frac{s(\theta)}{r^{\frac32}}\Big(\big(\partial_\theta\partial_r^{-1}\chi\big)^2+2\big(\partial_r^{-1}\chi\big)^2\Big)
+\frac18\int s(\theta) \cos^2\theta \Big(3r^{\frac12}+\frac{2\kappa}{r^{\frac12}}\Big)\chi^2\\ &~{}  +\int\frac1{r^{\frac12}} s(\theta) \chi \partial_{v}\chi|_{r=0}.
\end{aligned}
\end{equation}
The first two terms are nonnegative, but the last one is a flux through the horizon and has no definite sign. In fact, for $\kappa>0$ the horizon is non-degenerate and the red-shift effect damps the transverse derivatives at $r=0$, so no Aretakis instability is expected in the near-NHEK background.

The case $\kappa=0$ requires another argument. Indeed, it is not difficult to check that 
\[
\begin{aligned}
\frac{d}{dv} \frac12\int \frac1{r^{2}} s(\theta) \chi^2  \geq  &~{} \frac12\int \frac1{r^2}  s(\theta)\cos^2\theta\left( \partial_r^{-1} \partial_\theta \left( \frac{\chi}{\cos\theta}\right)\right)^2 \geq 0, 
\end{aligned}
\]
and therefore $\int \frac1{r^{2}} s(\theta) \chi^2(v)$ is increasing in $v$, under $\chi(v,0,\theta)\equiv0$. We call it a mild instability, in the sense that this local norm grows, but it might be the case that it finally converges to a final state.  In the general case where $\kappa=0$ but $\chi$ depends on $\theta$, if $\int_v \int \frac1{r^2}  s(\theta)\cos^2\theta\left( \partial_r^{-1} \partial_\theta \left( \frac{\chi}{\cos\theta}\right)\right)^2$ does not integrate in time $v\in [0,\infty)$ (notice that this might be case even with the integrand converging to zero), there will be a sort of averaged instability for non finite energy solutions.

\subsection{On Aretakis instability}

The Aretakis instability \cite{Aretakis,Aretakis:2012bm,Aretakis:2011hc} concerns in this case the behavior of the transverse derivatives $\partial_r^k\chi$ on the horizon, as opposite to the averaged quantities found in the previous paragraph. By applying $\partial_r^k$ to \eqref{eq:Aretakis_1} and restricting to the horizon $r=0$ gives, for each admissible mode $\chi_l$,
\[
\begin{aligned}
&2\,\partial_v\chi_l^{(k+1)}+2\kappa(k+1)\,\chi_l^{(k+1)}=\big[l(l+1)-k(k+1)\big]\,\chi_l^{(k)},
\\ 
& \chi_l^{(k)}:=\partial_r^k\chi_l\big|_{r=0},\quad k\ge0 .
\end{aligned}
\]
For $\kappa>0$ each $\chi_l^{(k+1)}$ satisfies a linear ODE in $v$ with damping rate $\kappa(k+1)$, so all transverse derivatives on the horizon decay once $\chi_l|_{r=0}$ does (redshift). There is no Aretakis instability on near-NHEK, as expected. For $\kappa=0$ the right-hand side vanishes at $k=l$, so
\[
H_l[\chi]:=\partial_r^{l+1}\chi_l|_{r=0} =\chi_l^{(l+1)}
\]
 is conserved along the Poincaré horizon $(r=0,v)$, and at $k=l+1$ one gets $\partial_v\chi_l^{(l+2)}=-(l+1)H_l$, i.e.\
\[
\partial_r^{l+2}\chi_l|_{r=0}(v)=\partial_r^{l+2}\chi_l|_{r=0}(0)-(l+1)H_l\,v,
\]
which grows linearly for every datum with $H_l\ne0$. This is the Aretakis mechanism \cite{Aretakis,Aretakis:2012bm} for the
gauge-invariant variable $\chi$ on NHEK. Notice that it is compatible with the decay statements of previous sections, which control weighted norms away from $r=0$ only.

\section{Symmetries and conserved quantities}\label{sec:charges}
In this section, we return to the unperturbed metric~\eqref{4dmetric}. To investigate the role of the symmetries as large gauge transformations, 
we first note that the spacetimes under consideration depend explicitly on 
time and are therefore non-stationary. We analyze the problem within the 
Hamiltonian framework of Regge and Teitelboim \cite{Regge:1974zd}, taking 
the radial coordinate as the evolution parameter. As we will discuss later on, this choice is natural 
for studying the long-time behaviour of the system and will allow us to 
connect the present results with the stability analysis of 
Section~\ref{sec:perturbations}. Most of the near-horizon charge literature is phrased in the covariant phase
space language of Iyer and Wald rather than in the Hamiltonian one; the
relation between the two formalisms is reviewed in \cite{Seraj:2016cym}, and
we appeal to it when contrasting our results with the covariant computations
of \cite{Castro:2021csm,Compere:2015bea} below.

The starting point is to write the metric in ADM form,
\begin{equation}\label{admetric}
ds^2 = N^2 dr^2 + h_{ij}\left(dx^i + N^i dr\right)
\left(dx^j + N^j dr\right),
\end{equation}
where $r$ is the radial coordinate, $h_{ij}$ is the induced metric 
on the three-dimensional timelike hypersurfaces $\Sigma$ defined by 
constant slices of $r$, and $N$ and $N^i$ are the lapse and shift 
functions, respectively. Latin indices label the coordinates of $h_{ij}$, 
and $\partial\Sigma$ denotes the boundary of $\Sigma$. The Hamiltonian is
\begin{equation}\label{radial-Hamiltonian}
H[N,N^{i}] = \int_{\Sigma} d^{3}x 
\left(N \mathcal{H} + N^{i}\mathcal{H}_{i}\right) + B[N,N^{i}],
\end{equation}
where $\mathcal{H}$ and $\mathcal{H}_i$ are the Hamiltonian and momentum 
constraints, which are proportional to Einstein's equations and therefore 
vanish on the near-NHEK solutions. The boundary term $B$ is added to 
ensure a well-posed variational problem.

The canonical generator of asymptotic symmetries is defined as 
\cite{Regge:1974zd}
\begin{equation}
Q[\xi,\xi^{i}] = B[\xi,\xi^{i}],
\end{equation}
where $(\xi, \xi^i)$ are the gauge parameters associated to the asymptotic 
Killing vector $\eta^\mu$ via the ADM decomposition,
\begin{equation}\label{defformation-vectors}
\xi = N\eta^{r}, \qquad \xi^{i} = \eta^{i} + N^{i}\eta^{r}.
\end{equation}
Its variation takes the form
\begin{equation}\label{eq:cargas}
\delta Q[\xi,\xi^{i}] = \int d^{2}S_{l}
\left[G^{ijkl}\left(\xi\nabla_{k}\delta h_{ij} - 
\delta h_{ij}\nabla_k\xi\right) + 2\xi_{k}\delta\pi^{kl} + 
\left(2\xi^{k}\pi^{jl} - \xi^{l}\pi^{kj}\right)\delta h_{jk}\right],
\end{equation}
with $G^{ijkl} = \frac{1}{2}\sqrt{h}
\left(h^{ik}h^{jl} + h^{il}h^{jk} - 2h^{ij}h^{kl}\right)$. 
Here $\pi^{ij}$ is the momentum conjugate to $h_{ij}$,
\begin{equation}
\pi^{ij} = \frac{\sqrt{h}}{2\kappa_{\rm g}}
\left(K^{ij} - h^{ij}K\right), \qquad 
K_{ij} = \frac{1}{2N}
\left(\nabla_{i}N_{j} + \nabla_{j}N_{i} - \partial_{r}h_{ij}\right),
\end{equation}
where $K_{ij}$ is the extrinsic curvature of $\Sigma$ and 
$\kappa_{\rm g} = 8\pi G$.

To carry over these expressions from the standard temporal foliation to the radial one,  where $h_{ij}$ 
carries Lorentzian signature $(-,+,+)$, we use the analytic continuation 
proposed in \cite{Banados:2004zt}: $N \to iN$, which induces 
$K \to -iK$ and $\sqrt{h} \to i\sqrt{h}$.

Finally, we recall that $H[\xi,\xi^i]$ is the generator of the gauge 
transformations \eqref{gauge-transformation}. Specifically, the field 
variations are obtained via Poisson brackets with the radial Hamiltonian,
\begin{equation}\label{Hamiltonian-symmetry-trans}
\delta h_{ij} = \left\{h_{ij}(x),\, H[\xi,\xi^{i}]\right\}, \qquad 
\delta\pi^{ij} = \left\{\pi^{ij}(x),\, H[\xi,\xi^{i}]\right\},
\end{equation}
which coincide with the Lie derivatives of the fields when evaluated 
on-shell \cite{Regge:1974zd,Banados:2004zt}.
When the foliation of the volume spacetime $\Sigma$ is set-up so that the leaves of the ADM foliation are timelike,  the conserved charges associated to the transformations laws \eqref{Hamiltonian-symmetry-trans}, should be constructed,  considering the radial evolution as the gravitational Poisson bracket,  which should occur for different times $t$ and $t'$ with fixed radius.  
\subsection{Radial ADM with Regge–Teitelboim boundary terms, charges at a late-time corner of a warped throat} \label{Corner-charges}

To define the conserved charges associated to the asymptotic symmetries, 
one must specify the physical boundary on which they are evaluated.  The surface loci should be defined for some asymptotic value of $t$, which as we will see later, define a natural physical boundary compatible with the chosen radial foliation of the volume spacetime $\Sigma$. In this setup, the conventional ADM picture is reversed and leaves of the foliation are timelike, and the ``evolution'' occurs along a spacelike direction.  Radial foliations of this type are standard in asymptotically AdS holography \cite{Regge:1974zd,Banados:2004zt} and have been employed in the near-horizon $\adstwo$ setting \cite{Castro:2019crn,Bredberg:2009pv}; here we adapt them to the warped geometry of the extremal Kerr throat. In this case, we deal with the absence of a Fefferman–Graham expansion by defining the charges directly by Regge–Teitelboim variation at a late-time cut.

Considering the Regge-Teitelboim expression \eqref{eq:cargas}, we write the the NHEK geometry in the ADM radial form,  and consider to express the symmetries in terms of the surface deformation vectors \eqref{general-Killin}, the variation of the surface charge is 
\begin{equation}
\delta Q[\epsilon ] |_{\partial \Sigma}= 8\pi c_{\phi} \delta M^{2}+4\pi\left(-\frac{\left(\frac{4 r_{0}^2}{\sqrt{\left(2 r_{0}^2+\mathcal{L}(t)\right)^2}}+1\right) \epsilon ''(t)}{r_{0}}+\epsilon (t) \left(2 r_{0}-\frac{\mathcal{L}(t)}{r_{0}}\right)\right)\delta M^{2},\label{eq:cargas-evaluadas}
\end{equation}
where $r_{0}$ highlights the fact that these charges are computed at constant radial hypersurfaces, and depend explicitly on $(r_{0},t)$.

The charges are evaluated on the two-surface
$\Sigma_{r_{0},t_{1}}=\{r=r_{0},\,t=t_{1}\}$, coordinatized by $(\theta,\phi)$
and the boundary of a leaf of the radial foliation, and we then let
$t_{1}\to+\infty$ at fixed $r_{0}$; we write $i^{+}$ for the resulting cut.
 We show that such
limit exists for a class of backgrounds considered down below, so that the explicit
$t$-dependence of \eqref{eq:cargas-evaluadas} enters through
$\epsilon(t)$ and $\mathcal{L}(t)$ in a finite  manner when evaluated at $i^{+}$.

\begin{figure}[t]
\centering
\begin{tikzpicture}[scale=0.9]
  \draw[thick] (0,-3.8) -- (0,3.8);
  \draw[thick] (3.1416,-3.8) -- (3.1416,3.8);
  \fill[gray!12] (3.1416,3.1416) -- (3.1416,-3.1416) -- (0,0) -- cycle;
  \draw[thick] (3.1416,3.1416) -- (0,0);
  \draw[thick] (3.1416,-3.1416) -- (0,0);
  \draw[very thick] (3.1416,-3.1416) -- (3.1416,3.1416);
  \draw[thick, densely dashed] plot[smooth] coordinates {
    (3.1416,-3.1416) (3.1062,-2.6490) (3.0200,-2.1998) (2.8372,-1.5258) (2.6643,-0.8721)
    (2.5587,0) (2.6643,0.8721) (2.8372,1.5258) (3.0200,2.1998) (3.1062,2.6490) (3.1416,3.1416)};
  \node[fill=gray!12, inner sep=1pt] at (2.85,0.24) {\scriptsize $Z{=}0.3$};
  \draw[thick] plot[smooth] coordinates {
    (3.1416,-3.1416) (3.0172,-2.6224) (2.9229,-2.4329) (2.6780,-2.0344) (2.0345,-1.1071) (1.6952,-0.5192)
    (1.5708,0) (1.6952,0.5192) (2.0345,1.1071) (2.6780,2.0344) (2.9229,2.4329) (3.0172,2.6224) (3.1416,3.1416)};
  \node[fill=gray!12, inner sep=1pt] at (1.99,0.24) {\scriptsize $Z{=}1$};
  \draw[thick, densely dotted] plot[smooth] coordinates {
    (3.1416,-3.1416) (3.0585,-2.9409) (2.8966,-2.7430) (2.4469,-2.2655) (1.6705,-1.4711) (1.2121,-1.0023)
    (0.8961,-0.6747) (0.5881,-0.3393) (0.4637,-0.1799)
    (0.3948,0) (0.4637,0.1799) (0.5881,0.3393) (0.8961,0.6747) (1.2121,1.0023) (1.6705,1.4711)
    (2.4469,2.2655) (2.8966,2.7430) (3.0585,2.9409) (3.1416,3.1416)};
  \node[fill=gray!12, inner sep=1pt] at (0.90,0.24) {\scriptsize $Z{=}5$};
  \fill (0,0) circle (2pt);
  \fill (3.1416,3.1416) circle (2pt) node[right] {\small $i^+$};
  \fill (3.1416,-3.1416) circle (2pt) node[right] {\small $i^-$};
\end{tikzpicture}
\caption{The figure shows the
constant-$Z$ ($R=1/Z$) worldlines of this diagram; each point of the picture is now a
squashed $(\theta,\phi)$ sphere rather than a spacetime point, and the frame
dragging is encoded in the twisting of its $\Uone$ fiber over the base, through
the connection $r\,dt$. Worldlines of constant $Z$ (static Poincar\'e observers), shown for $Z= 0.3$ (dashed), $Z = 1$ (solid) and $Z = 5$ (dotted); each label sits just to the right of its curve. All curves interpolate between the two corners $i^\pm$ of the triangle; the maximal depth into the bulk is $\tilde\sigma_{\max} = 2\arctan Z$, reached at $\tau = 0$. As $Z \to \infty$ the worldline degenerates onto the Poincar\'e horizons.}
\label{fig:constz}
\end{figure}
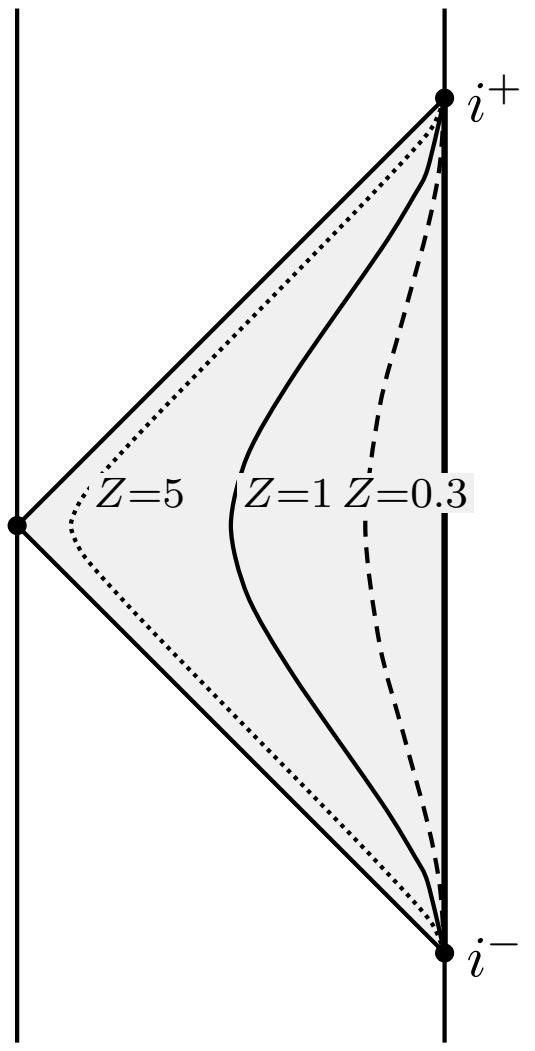

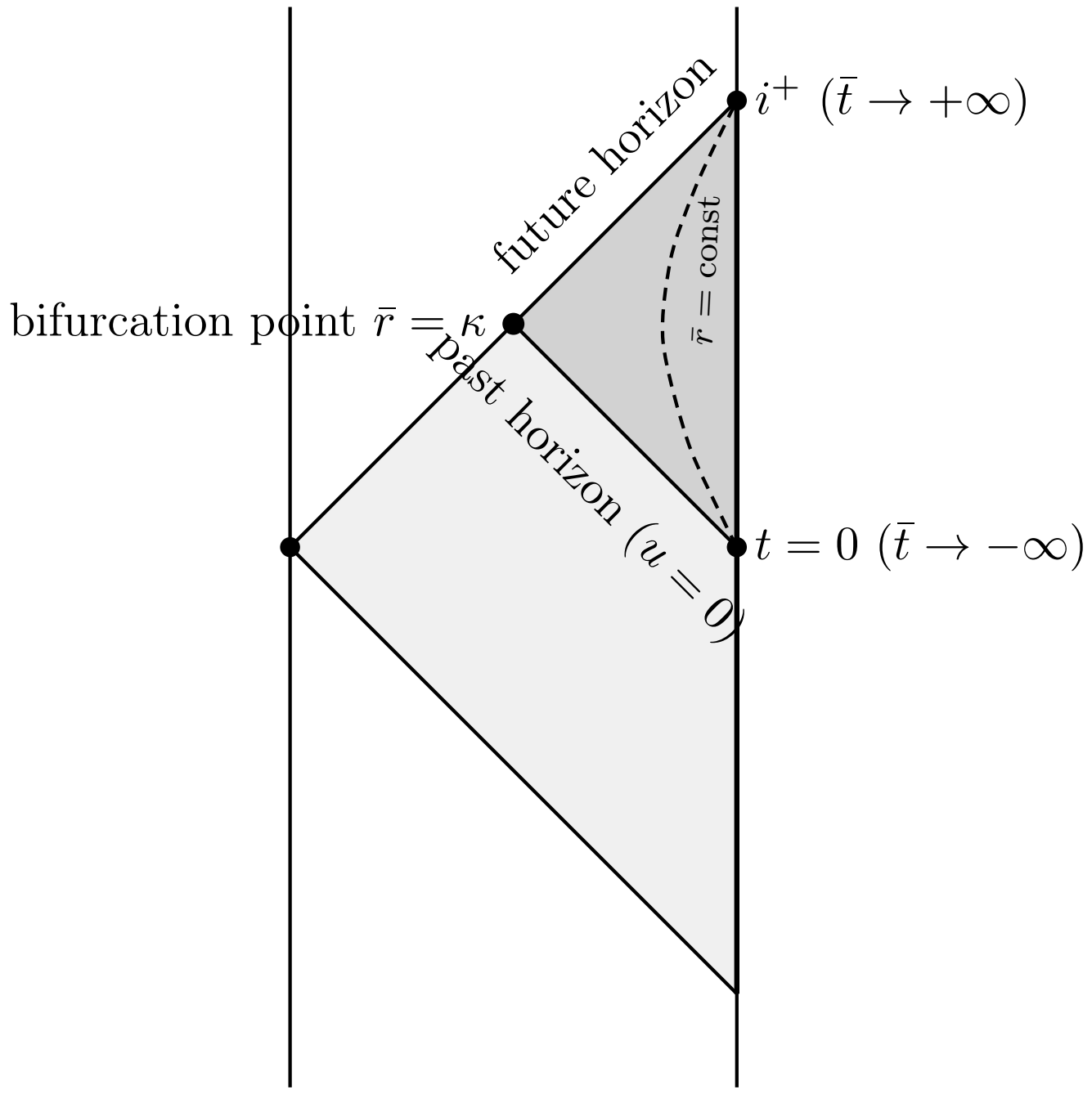
\begin{figure}[t]
\centering
\begin{tikzpicture}[scale=1.15]
  \draw[thick] (0,-3.8) -- (0,3.8);
  \draw[thick] (3.1416,-3.8) -- (3.1416,3.8);
  \fill[gray!12] (3.1416,3.1416) -- (3.1416,-3.1416) -- (0,0) -- cycle;
  \fill[gray!35] (3.1416,0) -- (3.1416,3.1416) -- (1.5708,1.5708) -- cycle;
  \draw[thick] (3.1416,3.1416) -- (0,0);
  \draw[thick] (3.1416,-3.1416) -- (0,0);
  \draw[thick] (3.1416,0) -- (1.5708,1.5708);
  \draw[very thick] (3.1416,-3.1416) -- (3.1416,3.1416);
  \draw[thick, densely dashed] plot[smooth] coordinates {
    (3.1416,0) (3.0427,0.19885) (2.7986,0.7378) (2.6224,1.4464) (2.6780,2.0344)
    (2.8431,2.5127) (3.0108,2.8776) (3.1416,3.1416)};
  \node[rotate=87] at (2.92,1.95) {\scriptsize $\bar{r}=\text{const}$};
  \node[rotate=45] at (2.20,2.72) {\small future horizon};
  \node[rotate=-45] at (2.10,0.42) {\small past horizon ($u=0$)};
  \fill (1.5708,1.5708) circle (2.2pt) node[left=2pt] {\small bifurcation point $\bar{r} =\kappa$};
  \fill (3.1416,0) circle (2pt) node[right] {\small $t=0$ ($\bar{t} \to-\infty$)};
  \fill (3.1416,3.1416) circle (2pt) node[right] {\small $i^+$ ($\bar{t}\to+\infty$)};
  \fill (0,0) circle (2pt);
\end{tikzpicture}
\caption{The black hole chart $(\bar{t},\bar{r})$ covers the dark wedge $0 < u < v$: the causal diamond of the boundary interval $T \in (0,\infty)$. Its Killing horizons meet at the bifurcation point, where $\bar{r} = \kappa$ at finite $\bar{t}$. A constant-$\bar{r}$ trajectory is shown dashed.}
\label{fig:wedge}
\end{figure}

The geometric picture behind this cut is the familiar one for the extremal
throat. Since the fiber block $d\theta^{2}+\Lambda^{2}(d\phi+r\,dt)^{2}$ of
\eqref{4dmetric} is positive definite, every causal curve projects to a causal
curve of the $\adstwo$ base and every causal base curve admits a
causal co-rotating lift, so the causal relations of (near-)NHEK are those of
$\adstwo$ and the conformal diagram of the base organizes the throat,
each of its points representing a squashed $(\theta,\phi)$ fiber
\cite{Spradlin:1999bn,Maldacena:1998uz}. In these terms NHEK is the Poincar\'e
triangle of Figure~\ref{fig:constz} and near-NHEK the sub-wedge of
Figure~\ref{fig:wedge}, related by the exponential map \eqref{eq:map1}; the
same pair of charts, and the same map between them, is used in
\cite{Hadar:2014dpa} to relate near-NHEK plunges to NHEK circular orbits.
The curves of fixed $r_{0}$ run between the corners $i^{\pm}$ of the diagram,
which is the sense in which the $t_{1}\to+\infty$ cut is a corner of the
throat. Details of the construction are collected in
Appendix~\ref{app:Conformal diagrams}.

Let us recall that the KdV equation constrains the Schwarzian mode
and emerges as a condition from the Killing equation \eqref{transformation}-\eqref{Killing-condition}.
In particular, it appears under the choice of symmetry generator
\begin{equation}\label{sym-gen-kdv}
\epsilon(t)=\epsilon_1(t) = \mathcal{L}(t)+ \bar{\epsilon}_1,
\end{equation}
where $\bar{\epsilon}_1$ is the integration constant of
\eqref{GD-polynomials}, which labels the class of admissible backgrounds
rather than a particular solution within it.

One relevant solution of \eqref{KdV} is the soliton \eqref{soliton-solution},
obtained for $\bar{\epsilon}_{1}=-\omega^{2}$, which satisfies
$\mathcal{L}(\pm\infty)=0$ and therefore approaches the NHEK geometry. It
depends on a single free parameter $\omega$, which fixes the geometry of the
solution in two correlated ways. The peak value is $\mathcal{L}(t_0)=\omega^2$
and, since $\mathcal{L}=\{f,t\}$ measures the deviation from pure NHEK, larger
$\omega$ corresponds to a stronger departure from extremality at the peak. The
characteristic temporal width is $\Delta t\sim 1/\omega$, as follows from the
argument of the $\cosh$. Larger $\omega$ therefore produces a narrow, sharply
peaked pulse, so that the deviation from extremality is intense but brief,
while small $\omega$ produces a broad, slowly varying profile in which the
deviation is mild but persists over a long time. Amplitude and width are not
independent, as is standard for KdV solitons.

The other case corresponds to the solution
\begin{equation}\label{soliton-with-background}
\mathcal{L}(t)=\frac{\omega^{2}}{\cosh^{2}\!\left(\frac{\omega(t-t_{0})}{2}\right)}
-\frac{\kappa^{2}}{2},
\end{equation}
which solves \eqref{KdV} provided
\begin{equation}\label{kappa-ebar}
\bar{\epsilon}_{1}=\tfrac{3}{2}\kappa^{2}-\omega^{2},
\qquad\text{equivalently}\qquad
\kappa^{2}=\tfrac{2}{3}\left(\bar{\epsilon}_{1}+\omega^{2}\right),
\end{equation}
with $\omega>0$. It satisfies $\mathcal{L}(\pm\infty)=-\kappa^{2}/2$, which by
\eqref{4near-NHEK} is the near-NHEK value of the Schwarzian mode, so that the
solution is asymptotic to the near-NHEK geometry and $\kappa$ is the same
parameter appearing there. The case $\bar{\epsilon}_{1}=-\omega^{2}$ gives
$\kappa=0$ and retrieves the NHEK scenario. The solution
\eqref{soliton-with-background} thus corresponds to shifting the soliton by a
constant background value, so that the geometry asymptotes to Rindler-$\adstwo$
rather than to pure NHEK, and its peak value is
$\mathcal{L}(t_0)=\omega^{2}-\kappa^{2}/2$.

In both cases the soliton is a transient deformation: the profile departs from
its asymptotic value over a time $\Delta t\sim1/\omega$ and returns to the same
value, approaching NHEK and near-NHEK respectively. The two solutions render
two different asymptotic values of the symmetry generator \eqref{sym-gen-kdv}
as $t\to+\infty$,
\begin{eqnarray}
&&\text{NHEK}\qquad\epsilon(+\infty)=-\omega^{2},\label{gauge-nhek}\\
&&\text{near-NHEK}\qquad\epsilon(+\infty)=\kappa^{2}-\omega^{2}.\label{gauge-near}
\end{eqnarray}
The general form of the charge variation \eqref{eq:cargas-evaluadas} at the
late-time cut, containing both cases, is
\begin{equation}\label{eq:cargas-asintoticas}
\delta Q|_{i^+}= 8\pi c_{\phi}\,\delta M^{2}
+4\pi\left[\epsilon(+\infty)\left(2 r_{0}
-\frac{\mathcal{L}(+\infty)}{r_{0}}\right)\right]\delta M^{2},
\end{equation}
since the term proportional to $\epsilon''(t)=\mathcal{L}''(t)$ in
\eqref{eq:cargas-evaluadas} decays exponentially on both solutions.

For the soliton asymptotic to NHEK the generator is \eqref{gauge-nhek} and
$\mathcal{L}(+\infty)=0$, so that the only contribution is
\begin{eqnarray}\label{NHEK-charge}
\text{NHEK}\qquad Q|_{i^+}&=& 8\pi c_{\phi} J-8\pi\left(\omega^{2} r_{0}\right)M^{2}.
\end{eqnarray}
We have imposed the extremality relation $J=M^{2}$ to associate each generator
with its corresponding conserved charge. The parameter $\omega$ controls a
transient pulse that decays away as $t\to\pm\infty$ on the Schwarzian
$\mathcal{L}$, but not on the gauge parameter \eqref{gauge-nhek}, which
survives at the cut and is what the charge records.\footnote{The generator of
time translations is $\epsilon(t)=c_{+1}$, associated to the
$\SLtwoR$ symmetry \eqref{SL(2,R)-generators}, the remaining ones
diverging at the proposed hypersurface. Together with the $\Uone$ generator,
\begin{equation}
Q|_{i^+} = 8\pi c_{\phi} J+2\pi \bar{c}_{+1} M^{2}.\label{eq:Charge SL2R}
\end{equation}}

In the second case, where the soliton asymptotes to the near-NHEK geometry,
the solution is governed by two independent scales. The background
$\kappa^{2}/2$ encodes a permanent deviation from extremality, so that the
black hole does not return to zero temperature but asymptotes to a finite
Hawking temperature $T\sim\kappa$, and there is an additional contribution
coming from the peak value of the soliton. With $\epsilon(+\infty)$ given by
\eqref{gauge-near} and $\mathcal{L}(+\infty)=-\kappa^{2}/2$,
\begin{equation}\label{near-NHEK-charge}
\text{near-NHEK}\qquad Q|_{i^+}= 8\pi c_{\phi} J
+4\pi\left(\kappa^{2}-\omega^{2}\right)
\left(2 r_{0}+\frac{\kappa^{2}}{2r_{0}}\right)M^{2},
\end{equation}
Throughout this section we hold $\omega$ and $\kappa$ fixed. They are not
gauge parameters but coordinates on the space of solutions, so that
\eqref{NHEK-charge} and \eqref{near-NHEK-charge} are the integrals of
\eqref{eq:cargas-asintoticas} along the $\delta M^{2}$ direction alone, where
integrability is immediate. Whether the charge remains integrable once
$\delta\omega$ and $\delta\kappa$ are included we leave open; the generator
\eqref{sym-gen-kdv} is field dependent, which is the situation in which
obstructions are known to arise \cite{Barnich:2010eb}.

The charge is well defined at any fixed value of the radial coordinate, but
\eqref{NHEK-charge} and \eqref{near-NHEK-charge} diverge as
$r_{0}\to\infty$ because of the term linear in $r_{0}$, and
\eqref{near-NHEK-charge} diverges as $r_{0}\to 0$ as well. One manner to
regularize the first divergence is to take
$\epsilon_{1}(+\infty)=\bar{\epsilon}_{1}^{\,2}/r_{0}$, which in practice means
reinterpreting the peak value of the soliton and the temperature as correlated
with the distance to infinity. With that choice \eqref{NHEK-charge} becomes
\begin{equation}
Q|_{i^+} = 8\pi c_{\phi} J+4\pi\,\bar{\epsilon}_{1}^{\,2}
\left(2-\frac{\mathcal{L}(+\infty)}{r_{0}^{2}}\right)M^{2},
\end{equation}
which converges as $r_{0}\to\infty$, since $\mathcal{L}(+\infty)$ vanishes for
the soliton without background and is finite in the near-NHEK case.

Alternatively, one may note that the radial dependence of
\eqref{eq:cargas-asintoticas} is stationary at a radius fixed by the geometry.
Writing $F(r_{0})=2r_{0}-\mathcal{L}(+\infty)/r_{0}$, one has
\begin{equation}
\frac{dF}{dr_{0}}=\frac{2r_{0}^{2}+\mathcal{L}(+\infty)}{r_{0}^{2}},
\end{equation}
which vanishes exactly at the horizon radius $r_{0}=\kappa/2$ identified in
Section~\ref{Extensions}, where $F=2\kappa$. Evaluated there, the near-NHEK
charge takes the finite value
\begin{equation}\label{near-NHEK-charge-horizon}
Q|_{i^+,\,r_{0}=\kappa/2}= 8\pi c_{\phi} J
+8\pi\kappa\left(\kappa^{2}-\omega^{2}\right)M^{2},
\end{equation}
with no further prescription. This choice degenerates in the NHEK case, where
$\mathcal{L}(+\infty)=0$ and the stationary radius collapses to $r_{0}=0$.

It is instructive to contrast \eqref{NHEK-charge} and \eqref{near-NHEK-charge}
with the charges of the Near-Horizon Extremal Geometries phase space of
\cite{Compere:2015mza,Compere:2015bea}, where the throat is deformed along the
torus of $\Uone$ isometry directions. Those deformations commute with the time
translation and scaling isometries of the background, so their symplectic
charges are independent of the surface on which they are evaluated and the
symmetry algebra is an extension of the Virasoro algebra with the entropy as
central term. Our deformation acts along time and breaks precisely those
isometries, which is why \eqref{NHEK-charge} and \eqref{near-NHEK-charge}
depend on $r_{0}$ and on the cut. For the same reason the $\SLtwoR$
charges, which vanish identically on the phase space of
\cite{Compere:2015mza} at fixed angular momenta and entropy, need not vanish
on the family \eqref{4dmetric}. A direct comparison does remain in the $\Uone$
sector where the $\partial_{\phi}$ charge of \cite{Castro:2021csm} is likewise
proportional to $J$. In this case, nonetheless, they compute charges of the
linear perturbation while here we restrain to charges of the background.

\section{Conclusions}\label{sec:conclusions}

We have studied axisymmetric gravitational perturbations of the near-horizon
region of (near-)extremal Kerr deformed by a Schwarzian mode, and two
structures organize the results.

The first is kinematic. Allowing the gauge parameter to depend on the field,
$\epsilon=\epsilon[\mathcal{L}]$, and requiring it to remain local, turns the
Killing equation of the family \eqref{4dmetric} into the stationary sector of
the Korteweg--de Vries hierarchy for $\mathcal{L}$. The first level gives NHEK
and near-NHEK ($\mathcal{L}'=0$); the second gives the stationary KdV equation
\eqref{KdV}, whose solitons are pulses of amplitude $\omega^{2}$ that depart
from and return to $\mathcal{L}(\pm\infty)=0$ (NHEK) or $-\kappa^{2}/2$
(near-NHEK). This is a selection of backgrounds rather than an evolution: the
hierarchy imposes constraints instead of generating a flow, and every member of
the family keeps the isometry algebra $\sltwoR\oplus\uone$. What the hierarchy
singles out is the subset of profiles on which one of the three generators is a
local functional of $\mathcal{L}$.

The second structure is dynamical. The linearized Einstein equations in the
axisymmetric sector reduce to a single gauge-invariant scalar $\chi$ obeying a
wave equation on the nearly-$\adstwo$ base, with $\mathcal{L}(t)$ as its only
background-dependent coefficient; for constant $\mathcal{L}$ this is the radial
Teukolsky equation, recovering \cite{Castro:2021csm}. Using energy and virial
methods, we found that the stability statement tracks the monotonicity of the
background. For $\mathcal{L}\equiv0$ the energy \eqref{E1-psi} is conserved and
the integrated local energy estimate \eqref{virial} gives asymptotic stability
in a weighted norm \eqref{virial_final}. On near-NHEK the conserved energy
associated with $\epsilon=\cosh(\kappa t)$ yields exponential decay of the norm
at a rate set by $\kappa$. For general $\mathcal{L}\ge0$ time-translation
invariance is lost, but \eqref{E_new} remains a Lyapunov functional,
$\dot E=\mathcal{L}'\times(\text{positive density})$: profiles with
$\mathcal{L}\ge\mathcal{L}_0>0$ and $\mathcal{L}'\in L^{1}$ give the uniform
bound $E(t)\le E_0e^{C_L/\mathcal{L}_0}$, while along the KdV soliton, where the
background weight is strong, only the part of the energy density supported
where $\mathcal{L}\gtrsim r^{2}$ relaxes, and only along a sequence of times.
In horizon-regular coordinates, the degenerate horizon of NHEK supports an
Aretakis-type instability (the $(l{+}1)$-th transverse derivative of each mode
is conserved and the $(l{+}2)$-th grows linearly in $v$), whereas on near-NHEK
the redshift damps all transverse derivatives. Finally, with a radial
Regge--Teitelboim foliation adapted to the non-stationary backgrounds, the
surface charges at a late-time corner of the throat distinguish the two
solitonic families through the pulse amplitude $\omega$ and the residual
temperature $\kappa$.

The picture that emerges is that ``stability of the near-horizon geometry'' is
not a single property. On NHEK, decay in a weighted energy norm coexists with
growth of transverse derivatives on the horizon; moving along the KdV family to
near-NHEK removes the latter and turns the conservation law for $\epsilon=1$
into a one-sided inequality, without any change in the local geometry. This
bears on the question of \cite{Durkee:2010ea}, whether an instability of the
near-horizon geometry implies one of the parent black hole: our results
indicate that it is underdetermined as posed, since the answer depends on the
norm and on the profile of the Schwarzian mode, neither of which is fixed by
the near-horizon limit. This complements \cite{Gralla:2016sxp,Gralla:2018xzo},
where the transient growth at finite temperature and the decay rates and
transverse-derivative growth follow from the near-horizon conformal symmetry
alone: within that symmetry, the Schwarzian profile is a further datum.

Several directions remain open. First, perturbations beyond the axisymmetric
sector: restricting to $m=0$ removes superradiance and makes the energy method
tractable, but for extremal Kerr the non-axisymmetric modes govern the
late-time behaviour and grow faster \cite{Casals:2016mel}. Second, the quantum
imprint of the background selection: the soft sector that labels these
backgrounds controls the low-temperature partition function of near-extremal
black holes \cite{Iliesiu:2020qvm}, and replacing Schwarzian by KdV asymptotics
already modifies the temperature dependence of the entropy in
Jackiw--Teitelboim gravity \cite{Cardenas:2024hah,Cardenas:2026}. Since the Schwarzian term
itself arises as a counterterm fixed by the parent theory
\cite{Castro:2025mnp}, computing the partition function over the family
\eqref{4dmetric}, level by level in the hierarchy, would show whether the
classical sensitivity to the profile survives quantization. Third, the
construction uses little beyond the $\adstwo$ factor and the warped fibration,
so extremal Kerr--Newman, Reissner--Nordstr\"om and higher-dimensional extremal
horizons should admit the same field-dependent symmetry analysis, indicating
how much of what we found is general to $\adstwo$ throats and how much is
specific to rotation.

\section*{Acknowledgements}

The authors received support from the Center for Mathematical Modeling (Universidad de Chile \& CNRS IRL 2807) through ANID/Basal project \# FB210005: H.VDB received partial funding from ANID/Fondecyt project \# 125-0596. This work is part of the CMM PRISMALab initiative. C.M. received funding via Fondecyt ANID 1231250.
M.C.\ is grateful to I.~J.~Araya and R.~Olea, for valuable
discussions.  Special thanks to  R.~Mancilla for many stimulating discussions and for
insights that significantly improved this work.

\appendix

\section{Teukolsky's master equation} \label{Apendice B}

An important feature of the Einstein equations that arises when studying gravitational perturbations is that those equations contain, for the geometric part, only the metric, the Ricci scalar $R$, and the Ricci tensor $R_{\mu\nu}$, where the latter one is a trace of the Riemann tensor ($R_{\mu\nu} = g^{\alpha\beta} R_{\alpha\mu\beta\nu}$). On the other hand, we have the Weyl tensor $C_{\mu\nu\alpha\beta}$, which is the traceless part of the Riemann tensor


\begin{equation}
C_{\mu\nu\alpha\beta} = R_{\mu\nu\alpha\beta} - 
g_{\mu[\alpha}R_{\beta]\nu} + g_{\nu[\alpha}R_{\beta]\mu} + 
\frac{1}{3}R g_{\mu[\alpha}g_{\beta]\nu}.
\end{equation}
Due to this, the Weyl tensor is left unconstrained by the Einstein equations and matter fields, and therefore represents the purely gravitational field. This is why the Weyl tensor has crucial information about the gravitational perturbations. With this tensor, one can also build the Weyl scalars: five scalars $\Psi_i$ that are invariant under diffeomorphisms,
\begin{align}
\Psi_0 &= C_{\mu\nu\alpha\beta} l^{\mu}m^{\nu}l^{\alpha}m^{\beta},\\
\Psi_1 &= C_{\mu\nu\alpha\beta} l^{\mu}n^{\nu}l^{\alpha}m^{\beta},\\
\Psi_2 &= C_{\mu\nu\alpha\beta} l^{\mu}m^{\nu}\bar{m}^{\alpha}n^{\beta},\\
\Psi_3 &= C_{\mu\nu\alpha\beta} l^{\mu}n^{\nu}\bar{m}^{\alpha}n^{\beta},\\
\Psi_4 &= C_{\mu\nu\alpha\beta} n^{\mu}\bar{m}^{\nu}n^{\alpha}\bar{m}^{\beta}.
\end{align}
Using these five scalars, it is possible to classify different spacetimes according to the value of said scalars. This classification is called the Petrov classification, and we will focus on Petrov type D spacetimes, which are characterized by having only 1 non-zero Weyl scalar \cite{chandrasekhar1998mathematical}: $\Psi_2$.

Keeping this idea in mind, it results natural to think of perturbations to the Weyl scalars $\Psi = \bar{\Psi} + \delta\Psi$, due to perturbations of the metric, but without the need to compute the perturbation $h_{\mu\nu}$. In fact, by studying $\delta\Psi$ one can reconstruct $h_{\mu\nu}$. This is how \cite{Teukolsky:1973ha} 
 studied the previous works on the Newman-Penrose formalism and found his master equation:
\begin{equation}
\begin{split}
& \qty[\frac{\qty(r^2 + a^2)^2}{\Delta} - a^2\sin^2\theta]\partial_t^2\Psi^{(s)} + 
\frac{4Mar}{\Delta}\partial_t\partial_\phi\Psi^{(s)} + 
\qty[\frac{a^2}{\Delta} - \frac{1}{\sin^2\theta}]\partial_\phi^2\Psi^{(s)}\\
& \quad - \Delta^{-s}\partial_r\qty(\Delta^{s + 1}\partial_r \Psi^{(s)}) - 
\frac{1}{\sin\theta}\partial_\theta\qty(\sin\theta \partial_\theta \Psi^{(s)}) - 
2s\qty[\frac{a(r - M)}{\Delta} + \frac{i\cos\theta}{\sin^2\theta}]\partial_\phi \Psi^{(s)}\\
& \quad - 2s\qty[\frac{M(r^2 - a^2)}{\Delta} - r - ia\cos\theta]\partial_t\Psi^{(s)} + 
(s^2\cot^2\theta - s)\Psi^{(s)} = 0.
\end{split}
\end{equation}
This master equation stands for any spin $s$ perturbation field $\Psi^{(s)}$, e. g., $\Psi^{(\pm 2)}$ for gravitational perturbations, $\Psi^{(\pm\frac{1}{2})}$ for fermionic perturbations, etc.  We will focus on perturbations to the Weyl scalars, with $s = \pm 2$. By performing the diffeomorphism \eqref{near_NHEK_coord}, for a field with no explicit dependence in $\phi$, the resulting equation adapted to the near-NHEK geometry is
\begin{equation}
\begin{split}
& \frac{16r^2}{\qty(4r^2 - \kappa^2)^2}\partial_t^2\Psi^{(s)} - \frac{M^4(4r^2 - \kappa^2)^2}{(4M^2 - \kappa^2)^2r^2}\partial_r^2\Psi^{(s)} - 
\frac{2M^2(1+s)(4r^2 + \kappa^2)}{(4M^2 - \kappa^2)r}\partial_r\Psi^{(s)}\\
& - \frac{1}{\sin\theta}\partial_\theta \qty(\sin\theta\partial_\theta\Psi^{(s)}) - 
\frac{8sr(4r^2 + \kappa^2)}{\qty(4r^2 - \kappa^2)^2}\partial_t\Psi^{(s)} + 
(s^2\csc^2\theta - s^2 - s)\Psi^{(s)} = 0.
\end{split}
\label{teu_near_NHEK}
\end{equation}
Performing the separation of variables
\begin{equation}
\Psi^{(\pm2)} = \sum_l U_l(t, r) S_l(\theta),
\end{equation}
the equation \eqref{teu_near_NHEK} splits into
\begin{equation}
S_l'' + \cot\theta S_l' = -l(l + 1)S_l,\\
\end{equation}
\begin{equation}
\begin{split}
-\frac{16r^2}{(4r^2 - \kappa^2)^2}\partial_t^2 U_l + \frac{M^4(4r^2 - \kappa^2)^2}{r^2(4M^2 - \kappa^2)^2}\partial_r^2 U_l + \frac{2M^2(1 + s)(4r^2 + \kappa^2)}{r(4M^2 - \kappa^2)}\partial_r U_l \\
+ \frac{8rs(4r^2 + \kappa^2)}{(4r^2 - \kappa^2)^2}\partial_t U_l
- (l(l + 1) - s - s^2 + s^2\csc^2\theta)U_l &= 0.
\end{split}
\end{equation}

\section{Proofs of energy estimates} \label{app:energy_estimates}

Proof of \eqref{cons_energy}. 
We multiply \eqref{eqn} by $s(\theta) \psi_t$ to obtain
\[
\begin{aligned}
\frac1{r^{4}} \psi_{tt}\psi_t s(\theta)  = &~{}  \psi_{rr} \psi_t s(\theta) \\
&~{} +  \frac1{r^2} \left( s(\theta) \psi_t  \psi_{\theta\theta} -3s(\theta) \cot (\theta) \psi_t  \psi_\theta -2s(\theta) \psi \psi_t \right).
\end{aligned}
\]
Therefore,
\[
\begin{aligned}
\frac{d}{dt} \left( \frac1{2r^{2}} \psi_t^2 s(\theta) \right)  = &~{}  s(\theta) r \psi_{rr} \psi_t  \\
&~{} -  \frac{2}{r} \psi_r  \psi_t s(\theta) +  \frac{d}{dt} \left(  \frac{\psi^2}{r^2} s(\theta) \right) +  r\psi_t s(\theta) \psi_{\theta\theta} -3r \sin^2 \theta \cos\theta \psi_\theta \psi_t  - \frac{d}{dt} \left(r \psi^2 s(\theta)     \right).
\end{aligned}
\]
Thus
\[
\begin{aligned}
\frac{d}{dt} \left( \frac1{2r^{4}} \psi_t^2 s(\theta) \right)  = &~{} \partial_r \left( \psi_{r} \psi_t s(\theta) \right) - \frac{d}{dt} \left( \frac12 \psi_{r}^2 s(\theta) \right) \\
& +\partial_\theta \left(  \frac1{r^2}s(\theta) \psi_{\theta}\psi_t  \right) - \frac{d}{dt} \left( \frac1{2r^2} s(\theta) \psi_\theta^2\right) \\
&- \frac{d}{dt} \left( \frac1{r^2} \psi^2 s(\theta) \right) .
\end{aligned}
\]
This gives \eqref{cons_energy}.

Proof of \eqref{cons_mom}. We compute and use \eqref{eqn}:
\[
\begin{aligned}
& \partial_t \left( \frac1{r^2}s(\theta)  \psi_r \psi_t \right) \\
&~{} =  \frac1{r^2}s(\theta)  \psi_{tr} \psi_t  +  \frac1{r^2}s(\theta)  \psi_r \psi_{tt}  \\
&~{} =  \partial_r \left(s(\theta) \frac{\psi_t^2}{2r^2} \right) +s(\theta) \frac{\psi_t^2}{r^3} +  s(\theta)  \psi_r \left( r^2\psi_{rr} + \psi_{\theta\theta} -3\cot\theta  \psi_\theta -2 \psi \right)  \\
 &~{} = \partial_r \left(s(\theta) \frac{\psi_t^2}{2r^2} \right) +s(\theta) \frac{\psi_t^2}{r^3} +\frac12s(\theta) \partial_r (r^2\psi_r^2) - s(\theta) r \psi_r^2 \\
&~{} \quad +\partial_\theta( s(\theta) \psi_\theta \psi_r) - \partial_r \left( s(\theta) \frac{\psi_\theta^2}2 \right)  -\partial_r( \sin^{-3} \theta \psi^2 ) . 
\end{aligned}
\]
This proves \eqref{cons_mom}.

\medskip

Proof of \eqref{cons_nada}. Simply compute and use \eqref{eqn}; one gets
\[
\begin{aligned}
\partial_t \left( \frac1{r^3}s(\theta)  \psi \psi_t \right) =&~{}  \frac1{r^3}s(\theta)  \psi_t^2  +  \frac1{r^3}s(\theta)  \psi \psi_{tt}  \\
=&~{}  \frac1{r^3}s(\theta)  \psi_t^2   +  s(\theta)  \psi \left( r \psi_{rr} + \frac1r\psi_{\theta\theta} -3\frac1r\cot\theta  \psi_\theta -2\frac1r\psi \right)  \\
= &~{} s(\theta) \frac{\psi_t^2}{r^3} +s(\theta) \left( \partial_r(r\psi\psi_r) -\psi\psi_r -r\psi_r^2  \right) \\
&~{} + \partial_\theta \left( \frac1r s(\theta) \psi\psi_\theta \right) - \frac1r s(\theta) \psi_\theta^2 - \frac2rs(\theta) \psi^2. 
\end{aligned}
\]
Proof of \eqref{virial}. First of all, from \eqref{cons_mom} and \eqref{cons_nada},
\[
\frac{d}{dt} \int s(\theta) \left( \frac1{r^2} \psi_r \psi_t  - \frac1{r^3}  \psi \psi_t \right) =  \int s(\theta) \left( \left( \frac12 +\frac1r\right) \psi_\theta^2 + \frac2r \psi^2 \right).
\]
Notice that 
\[
\left|  \int s(\theta) \left( \frac1{r^2} \psi_r \psi_t  \pm \frac1{r^3}  \psi \psi_t \right)  \right| \lesssim \int s(\theta) \left(\psi_r^2 +\frac{\psi_t^2}{r^4} + \frac{\psi^2}{r^2} \right)  \lesssim E.
\]
This proves that 
\begin{equation}\label{previous}
 \int_0^t \int_{r,\theta}  \frac{s(\theta)}{r} \left( \psi_\theta^2 + 2\psi^2 \right)  \leq C(E).
\end{equation}
To get \eqref{virial}, 
let $m=m(r)$ be a bounded smooth function, with $m(0)=0$. Again using \eqref{cons_mom} and \eqref{cons_nada},
\begin{equation}\label{cons_mom_2}
\begin{aligned}
\frac{d}{dt}  \int m(r) s(\theta)\left( \frac{\psi_r \psi_t}{r^2} -\frac{\psi \psi_t}{r^3}  \right)
={}& - \frac12\int  m'(r)\, s(\theta)  \left( \frac{\psi_t^2}{r^2}   + r^2\psi_r^2 \right) \\
&   +\frac12  \int  \left( m'(r) +\frac{2m(r)}{r} \right)s(\theta)\, \psi_\theta^2\\
&  +\int \left( \frac{2m(r)}r -\frac12m''(r)r \right) s(\theta)\,\psi^2 .
\end{aligned}
\end{equation}

Taking $m(r)=\arctan r$, we get $| m'(r) +\frac{2 m(r)}r  | \lesssim 1 +\frac1r $, and $|\frac{2m(r)}r -\frac12m''(r)r |\lesssim \frac1r$, and integrating in \eqref{cons_mom_2} and using \eqref{previous},
\[
\int_0^t \int  \frac{ s(\theta) }{1+r^2} \left( \frac{\psi_t^2}{r^2}   + r^2\psi_r^2 \right) \lesssim E + \int_0^t \int s(\theta) \left( \left(1+\frac1r \right) \psi_\theta^2 + \frac1r\psi^2 \right)   \lesssim C(E).
\]
This proves \eqref{virial}. Consequently, there exists an increasing sequence of time $t_n\to +\infty$ such that 
\[
\begin{aligned}
& \lim_{n\to+\infty}  \int_{r,\theta} \frac{s(\theta)}{1+r^2} \left(  \left(\psi_\theta^2 + 2\psi^2  \right)  + r^2 \psi_r^2 +\frac{\psi_t^2}{r^2}  \right)(t_n)  \\
& \qquad \leq 2\lim_{n\to+\infty}  \int_{r,\theta} s(\theta) \left(  \left(1+\frac1r \right) \psi_\theta^2 + \frac1r\psi^2   + \frac{1}{1+r^2} \left( r^2 \psi_r^2 +\frac{\psi_t^2}{r^2} \right) \right)(t_n) =0.
\end{aligned}
\]
From \eqref{cons_energy}, we get
\[
\begin{aligned}
& \left| \int \frac{r^2\sin^{-3}\theta}{1+r^2} \left( \frac{\psi_t^2}{r^4} +\frac{2\psi^2}{r^2} +\psi_r^2 +\frac{\psi_\theta^2}{r^2}\right)(t) -\int  \frac{r^2 \sin^{-3}\theta }{1+r^2}\left( \frac{\psi_t^2}{r^4} +\frac{2\psi^2}{r^2} +\psi_r^2 +\frac{\psi_\theta^2}{r^2}\right)(t_n)\right| \\
& ~{} \leq \int_{t}^{t_n}\frac{r}{(1+r^2)^2} \int  \sin^{-3}\theta |\psi_{r} \psi_t| \lesssim \int_t^{t_n} \int  \frac{ \sin^{-3}\theta }{1+r^2} \left( \frac{\psi_t^2}{r^2}   + r^2\psi_r^2 \right) \leq C(E).
\end{aligned}
\]
Sending $n\to +\infty$ and then $t\to +\infty$, we get the desired result. This proves \eqref{virial_final}.

\section{The twisted fibration over \texorpdfstring{$\adstwo$}{AdS2} and its conformal diagrams} \label{app:Conformal diagrams}

In the near-horizon extreme Kerr geometry \eqref{NHEK}, the $(T,R)$ block is
Poincar\'e $\adstwo$ with $R = 1/Z$, and the full geometry is a twisted
$\Uone$ fibration over it, warped over $\theta$. Because the fiber block
$d\theta^2 + \Lambda^2(d\Phi + R\,dT)^2$ is positive definite, the projection of
any causal curve onto the $(T,R)$ base is causal in the $\adstwo$ metric,
and any causal base curve lifts to a causal NHEK curve (take $\dot\theta = 0$
and the co-rotating lift $\dot\Phi = -R\,\dot T$). The causal relation between
fibers of NHEK is therefore exactly the causal relation between points of
$\adstwo$, so the conformal diagram of the base organizes the throat,
each of its points representing a squashed, twisted $(\theta,\Phi)$ sphere.
This is what allows the two-dimensional diagrams used in
Section~\ref{Corner-charges} to describe a warped four-dimensional geometry.

The construction of those diagrams is standard
\cite{Spradlin:1999bn,Maldacena:1998uz}. In the Poincar\'e chart
$ds^2 = Z^{-2}(-dT^2 + dZ^2)$, with $T \in \mathbb{R}$ and $Z \in (0,\infty)$,
the null coordinates $u = T-Z$, $v = T+Z$ and the compactification
$U = \arctan u$, $V = \arctan v$ bring the metric to the global form
\begin{equation}\label{globalstrip}
ds^2 = \frac{1}{\sin^2\tilde\sigma}\left(-d\tau^2 + d\tilde\sigma^2\right),
\qquad \tau = U+V, \quad \tilde\sigma = V-U,
\end{equation}
while the patch condition $v - u = 2Z > 0$ becomes the triangle
$\tilde\sigma \in (0,\pi)$, $|\tau| < \pi - \tilde\sigma$ inside the global
strip. Its vertical base is a segment of one conformal boundary, carrying all
of Poincar\'e time through $T = \tan(\tau/2)$; its two null edges are the past
and future Poincar\'e horizons, which are the Killing horizons of $\partial_T$;
and its apex is the single point $(\tau,\tilde\sigma) = (0,\pi)$ at which the
patch touches the second boundary. The two base corners are $i^{\pm}$ and the
apex is $i^{0}$.

The curves relevant to the charges of Section~\ref{Corner-charges} are the
worldlines of constant $Z$, equivalently of constant $R$, which are the orbits
of $\partial_T$. Writing
\begin{equation}\label{transftoglobal}
\tau = \arctan(T+Z) + \arctan(T-Z), \qquad
\tilde\sigma = \arctan(T+Z) - \arctan(T-Z),
\end{equation}
every such curve runs from the bottom corner of the triangle at
$T \to -\infty$ to the top corner at $T \to +\infty$, whatever the value of
$Z$, and reaches its maximum depth $\tilde\sigma_{\max} = 2\arctan Z$ at
$\tau = 0$ (Fig.~\ref{fig:constz}). Small $Z$ hugs the boundary, $Z = 1$
reaches the middle of the strip, and as $Z \to \infty$ the curve degenerates
onto the pair of null horizons. This is the sense in which the late-time cut
used in Section~\ref{Corner-charges} is a corner of the diagram.

\subsection*{From NHEK to near-NHEK}
\label{sec:map1}

To describe near-NHEK in the same terms we use the coordinate transformation
\begin{equation}\label{eq:map1}
T = \frac{e^{\kappa t}\,\rho}{\kappa\sqrt{\rho^{2}-\kappa^{2}}}, \qquad
R = e^{-\kappa t}\sqrt{\rho^{2}-\kappa^{2}}, \qquad
\Phi = \phi + \frac{1}{2}\ln\frac{\rho - \kappa}{\rho + \kappa},
\end{equation}
whose inverse reads
\begin{equation}
T \pm Z = \frac{e^{\kappa t}}{\kappa}\sqrt{\frac{\rho \pm \kappa}{\rho \mp \kappa}},
\end{equation}
and which covers the sub-wedge $0 < u < v$ of the Poincar\'e patch, with
horizon at $\rho = \kappa$. Applied to \eqref{NHEK}, the full map
$(T,R,\theta,\Phi) \to (t,\rho,\theta,\phi)$ yields
\begin{equation}
ds^2 = \Lambda(\theta)\left[-(\rho^{2} - \kappa^{2})\,dt^{2}
+ \frac{d\rho^{2}}{\rho^{2} - \kappa^{2}} + d\theta^2\right]
+\Gamma(\theta)\left[ d\phi + \rho\, dt\right]^2,
\label{eq:nearnhek}
\end{equation}
which is \eqref{4near-NHEK} after the change
$r = \tfrac{1}{2}\left(\rho + \sqrt{\rho^{2}-\kappa^{2}}\right)$, equivalently
$\rho = r + \kappa^{2}/4r$. This is the same intermediate coordinate used in
Section~\ref{sec:aretakis}, and the horizon $\rho = \kappa$ corresponds to
$r = \kappa/2$, the radius identified in Section~\ref{Extensions}.

With the tortoise coordinate
$\rho_* = \frac{1}{2\kappa}\ln\frac{\rho - \kappa}{\rho + \kappa}$, which tends
to $-\infty$ as $\rho \to \kappa$, the null coordinates are
\begin{equation}\label{null-to}
u = \frac{1}{\kappa}\, e^{\kappa(t + \rho_*)},
\qquad
v = \frac{1}{\kappa}\, e^{\kappa(t - \rho_*)}.
\end{equation}
Sending $\rho \to \kappa$ at fixed $t$ gives $u \to 0$ and $v \to \infty$, so
that the horizon at finite time is the single point
\begin{equation}
(U, V) = \left(0, \tfrac{\pi}{2}\right),
\qquad \text{i.e.} \qquad
(\tau, \tilde\sigma) = \left(\tfrac{\pi}{2}, \tfrac{\pi}{2}\right),
\end{equation}
exactly where the bifurcation point sits in Figure~\ref{fig:wedge}; the two
full horizon branches are reached only as $t \to \pm\infty$ along the edges.
At fixed $\rho$ both $u$ and $v$ are proportional to $e^{\kappa t}$, so that
$v = \frac{\rho + \kappa}{\rho - \kappa}\,u$ is a straight ray through the
origin of the $(u,v)$ plane, whose image is the curve $\tan V = c\,\tan U$ with
$c = \frac{\rho + \kappa}{\rho - \kappa}$, drawn dashed in
Figure~\ref{fig:wedge} for $c = 3$, that is $\rho = 2\kappa$. Using
\eqref{transftoglobal} one checks that $t \to \infty$ at fixed $\rho$
approaches $\tilde\sigma = 0$ and $\tau = \pi$.

\end{document}